\documentclass[twocolumn,floats,pra,eqsecnum,showpacs]{revtex4-1}

\usepackage{amsmath}
\usepackage{amsfonts}

\usepackage{tikz}

\newcommand {\be}{\begin{equation}}
\newcommand {\ee} {\end{equation}}
\newcommand {\bea}{\begin{eqnarray}}
\newcommand {\eea} {\end{eqnarray}}
\newcommand{\non}{\nonumber}
\newcommand{\bk}{{\bf k}}

\newcommand{\bx}{{\bf x}}

\newcommand{\half}{{\textstyle\frac{1}{2}}}
\newcommand{\rt}{{\widetilde{r}}}
\newcommand{\qt}{{\widetilde{q}}}
\newcommand{\cT}{{\cal T}}
\newcommand{\eps}{{\varepsilon}}

\begin{document}

%@twocolumnfalse\endcsname

\title{One-step replica-symmetry-breaking phase below the de~Almeida--Thouless line\\in low-dimensional spin glasses}
\author{J. H\"oller$^1$ and N. Read$^{1,2}$}
\affiliation{$^1$Department of Physics, Yale University, P.O. Box 208120, New Haven, CT 06520-8120\\
$^2$Department of Applied Physics, Yale University, P.O. Box 208284, New Haven, CT 06520-8284}
\date{April 12, 2020}

\begin{abstract}
The de Almeida-Thouless (AT) line is the phase boundary in the temperature--magnetic field plane of an Ising spin glass at which a 
continuous (i.e.\ second-order) transition from a paramagnet to a replica-symmetry-breaking (RSB) phase occurs, according to 
mean-field theory. Here, using field-theoretic perturbative renormalization group methods on the Bray-Roberts reduced 
Landau-Ginzburg-type 
theory for a short-range Ising spin glass in space of dimension $d$, 
we show that at {\em nonzero} magnetic field the nature of the corresponding transition is modified as follows: a) for $d-6$ small and 
positive, with increasing field on the AT line, first, the ordered phase just below the transition becomes the so-called one-step RSB, 
 instead of the full RSB that 
occurs in mean-field theory; the transition on the AT line remains continuous with a diverging correlation length. Then at a higher field, 
a tricritical point separates the latter transition from a quasi-first-order one, that is one at which the correlation 
length does not diverge, and there is a jump in part of the order parameter, but no latent heat. The location of the tricritical point
tends to zero as $d\to6^+$; b)  for $d\leq 6$, we argue that the quasi-first-order transition could persist down 
to arbitrarily small nonzero fields, with a transition to full RSB still expected at lower temperature. Whenever the quasi-first-order 
transition occurs, it is at a higher temperature than the AT transition 
would be for the same field, preempting it as the temperature is lowered. These results may explain the reported absence of a diverging 
correlation length in the presence of a magnetic field in low-dimensional spin glasses in some simulations and in high-temperature series 
expansions.  
We also draw attention to the similarity of the ``dynamically-frozen'' state, which occurs at temperatures just above the 
quasi-first-order transition, 
and the ``metastate-average state'' of the one-step RSB phase, and discuss the issue of the number of pure states in either.
\end{abstract}

%\pacs{pacs} %]

\maketitle

%%%%%%%%%%%%%%%%%%%%%%%%%%%%%%%%%%%%%%%%%%%
\section{Introduction}
\label{intro}

\subsection{Background and motivation}

A transition in a classical Ising spin glass (SG) in a magnetic field within a mean-field treatment was found by de Almeida and 
Thouless (AT) \cite{at}, 
who showed that the mean-field solution found by Sherrington and Kirkpatrick (SK) \cite{sk} is unstable at sufficiently low 
temperature $T$ in any
magnetic field $h$, and so the instability or transition occurs on a line  (now known as the AT line) $T_{AT}(h)$ in the $T$--$h$ 
plane; the AT 
line passes through the critical temperature $T_c$ at $h=0$. In a short-range Ising SG (i.e.\ 
the Edwards-Anderson [EA] model \cite{ea}), at such a transition the correlation length and the SG susceptibility both diverge, typical of 
a continuous (or second order) phase transition. The AT instability indicated that in the SK model, or within mean field theory, the 
symmetry under permutations of the replicas (introduced by EA) must be broken in the phase below the AT line in nonzero as well as 
in zero magnetic field. The replica-symmetry breaking (RSB) ordering in the low-temperature phase was determined by Parisi 
 \cite{par79}, 
and has been proved to give the correct thermodynamic properties of the SK model \cite{guerra,tala}.  Among many reviews, the most 
relevant to this paper are the books, Refs.\ \cite{mpv_book,ddg_book}.

In the short-range models, a controversy has remained about whether the RSB picture is a correct description of the ordered phase 
in each dimension $d$ of space, at least for those $d$ in which a transition at nonzero temperature occurs at $h=0$. The leading 
 alternative
is the scaling-droplet picture \cite{bm1,macm,fh}, in which in particular there is no transition at nonzero $h$. Thus the question of the 
existence and nature of a transition in a magnetic field is important for our understanding of the SG ordered phase, especially in realistic 
 dimensions, 
say $d=3$. The problem has been studied in a number of simulations (in both the nearest-neighbor $d$-dimensional and one-dimensional 
 power law models; 
see for example Refs.\  \cite{janus0,janus1,janus2,lprr,lkmy}), and also using high-temperature series 
expansions \cite{singh}. Some of these works found no divergence of the correlation length in a magnetic field in low dimensions ($d<6$, 
and for corresponding power-laws in one dimension), though others did find such a divergence. 

The standard method of studying the effect of fluctuations around mean-field theory in short-range models is to use a statistical field
 theory 
with an action obtained from Landau-Ginzburg theory. Perhaps surprisingly, the analysis of the AT line (by which we will always mean at
 $h>0$; note 
that the sign of $h$ is immaterial for Ising spins) in the short-range case within such a treatment 
encounters difficulties in low dimensions ($d\leq 6$). In an important early paper, Bray and Roberts (BR) \cite{br} formulated a
 ``reduced'' action for the 
fluctuating modes (called ``replicons'') that remain massless on the AT line. They found that, at one-loop order, the perturbative
 renormalization group (RG) flows 
for the two coupling constants of this theory 
experience runaway flows to strong coupling for $d\leq 6$, so that no RG fixed point that could describe the behavior of the AT line for
 $d\leq6$ 
could be found within perturbation theory. They suggested that this might mean that either (i) the transition becomes first order (with no
 divergence 
of the correlation length), or (ii) the transition is first-order even in mean-field theory, or (iii) there is no transition in a nonzero magnetic
 field for $d\leq6$. 
The latter possibility has been used as an argument in favor of the scaling-droplet theory (see Ref.\ \cite{mb} and references therein,
 and also the 
response in Ref.\ \cite{pt}). It is also possible to imagine unusual non-perturbative scenarios with a second-order transition at $d\leq6$.
 In a later 
work \cite{cy}, it was proposed that an RG fixed point that arises in {\em two}-loop RG for the BR theory at $d<6$ could produce a
 second-order transition.
However, as such a fixed point can occur only at couplings of order one, the validity of the fixed point and its survival at higher order
 are not clear
(even if the theory happens to be Borel summable, as it was argued to be \cite{cy}). 
Later still, Moore and one of the authors (Ref.\ \cite{mr}; to be referred to as MR) showed that the one-loop BR flows also imply 
that there should be a multicritical point on the AT line as $d\to6^+$. Other authors have found evidence in hierarchical-lattice 
models (i.e.\ using Migdal-Kadanoff RG) that the transition in a magnetic field is controlled by a zero-temperature 
fixed point \cite{bm84,ab}; such models differ drastically from the EA model \cite{mr}. 

In general phase-transition theory, the possibility that a transition is first-order can rarely be ruled out entirely, and has frequently been 
stated to be a possible solution to the problem raised by BR. However,  it was pointed out long ago that, within a Landau theory of a SG 
formulated in terms of replicas, a conventional first-order transition with positive latent heat is not possible; it cannot originate from
 solving 
such a theory \cite{gks}. The reason is that in replica theory (including Parisi's RSB scheme), the free energy functional in the limit of
 $n=0$ replicas must 
be maximized with respect to Parisi's $q(x)$ function, not minimized, and so a crossing of the free energies of extrema of this functional
 would produce a latent heat 
that is either zero or negative; the latter is forbidden by conventional thermodynamics. 

However, there is a way to obtain a quasi-first-order transition (i.e.\ first order with zero latent heat) from a RSB Landau theory. It was
 identified 
by Gross, Kanter, and Sompolinsky (GKS) \cite{gks} when they were studying SGs of spins with either Potts 
or uniaxial quadrupolar symmetry in zero magnetic field. Assuming isotropic SG order (an assumption that need not concern us here), the
 SG order parameter 
becomes a matrix 
$Q_{\alpha\beta}$ ($\alpha$, $\beta=1$, \ldots, $n$) that is symmetric, with $Q_{\alpha\alpha}=0$ for all $\alpha$, as for the Ising
 case. Their
 Landau theory, that is the free energy expanded in powers of $Q_{\alpha\beta}$, was (with slight changes of notation for consistency
 with the present paper)
\bea
F&=&\frac{1}{4}\rt\sum_{\alpha,\beta}Q_{\alpha\beta}^2-\frac{1}{6}w_1\sum_{\alpha\beta\gamma}Q_{\alpha\beta}
  Q_{\beta\gamma}Q_{\gamma\alpha}
-\frac{1}{6}w_2\sum_{\alpha\beta}Q_{\alpha\beta}^3\non\\
&&{}-\frac{1}{8}y\sum_{\alpha\beta}Q_{\alpha\beta}^4;
\label{gks}
\eea
the terms up to the cubic order are the most general form allowed by the symmetry of the Potts and quadrupolar models. (The quartic
 term with coefficient $y$ 
is not the most general form; we come to that later.) Ising SGs in zero magnetic field are usually described by the same Landau theory,
 except that 
there $w_2=0$ as a consequence of inversion symmetry in spin space, and $y>0$. GKS found that for $\rt$ not too large and negative  
(i) for $y\leq 0$ and $0<w_2/w_1<1$, there is a continuous transition at $\rt=\rt_c=0$, but for $\rt<\rt_c$ the Parisi function $q(x)$ is
 a step function 
(of $x\in [0,1]$) instead of the continuous function familiar for the Ising spin glass in mean field theory, and (ii) for $y<0$ and $w_2
/w_1>1$, the transition 
is discontinuous: $q(x)$ is again a step function, but $q(1)$ has a jump at $\rt_c$, and $\rt_c$ is now positive; there is no latent heat.
 In case (ii), the 
eigenvalues of the Hessian are strictly positive as $\rt\to\rt_c$ on both sides of the transition, implying that the SG susceptibility and, in
 a finite-dimensional version, 
the correlation  length do not diverge at $\rt_c$. The step function form of $q(x)$ describes what is known as one-step RSB (or 
$1$-RSB), and the 
quasi-first-order transition in case (ii) has the form of the transition in the random energy model (REM) \cite{derr,gm}, though there 
the extensive part of the entropy is 
zero in the low-temperature region, which is not the case here. 

The non-derivative part of the BR reduced action has the same form as eq.\ (\ref{gks}) through terms of cubic order, except that it
 involves, in place of 
$Q_{\alpha\beta}$,  the field $\widetilde{Q}_{\alpha\beta}$ which satisfies the additional conditions $\sum_\alpha 
\widetilde{Q}_{\alpha\beta}=0$, that 
define the replicon subspace. (The terms through cubic order give the most general cubic action in the replicon sector.) $w_2$ can be
 nonzero, 
due to the breaking of inversion symmetry by the magnetic field. Moreover, the BR RG flows for $w_1$, $w_2$ imply that for $w_2\neq
 0$ the ratio 
$\rho=w_2/w_1$ tends to a value $\rho^*=14.379\ldots$ on the AT line for $d\leq 6$. As this is larger than unity, the GKS results could
 come into play. But 
then $y<0$ is also necessary. The initial value of $y$ in BR is positive, but the RG flows might take the parameters into a region where
 GKS can be applied. 
 Previous works do not seem to have considered the quartic terms that could be included in the BR action. Presumably this was because
 quartic and higher-order 
terms are irrelevant in the RG sense near $d=6$ dimensions. However, Fisher and Somplinsky (FS) \cite{fs} explained that, because the
 quartic terms cannot be 
dropped at and below a SG transition, they are ``dangerously irrelevant'', and moreover they are important for the scaling behavior
 when $d<8$, because 
of the form of their RG flows, even though they are irrelevant. For example, the form of the AT line at small $h$ depends on $y$, and so
 is modified for 
$6<d<8$, to interpolate from the mean-field results for $d>8$ to the scaling forms for $d\leq6$ (see also Ref.\ \cite{gmb}). 
The GKS results indicate that an extreme form of dangerous irrelevance could occur, because reversing the sign of $y$ causes {\em
 qualitative} changes in 
the phase transition behavior, not just quantitative changes such as in exponents. 

These considerations motivate us to consider a Landau-Ginzburg field theory that extends the BR reduced theory by including quartic
 terms. The fields in the theory are 
the same ones, $\widetilde{Q}_{\alpha\beta}(\bx)$, and the action is now
\bea
F[\{\widetilde{Q}_{\alpha\beta}\}]
&=& \int   d^dx\,   \left[{\textstyle\frac{1}{4}}\sum(\nabla
\widetilde{Q}_{\alpha\beta})^2+{\textstyle\frac{1}{4}}\widetilde{r}
\sum \widetilde{Q}_{\alpha\beta}^2
 \right.  \non \\
&&\left.{}-{\textstyle\frac{1}{6}}w_1
\sum \widetilde{Q}_{\alpha\beta}\widetilde{Q}_{\beta\gamma}\widetilde{Q}_{\gamma\alpha}
-{\textstyle\frac{1}{6}}w_2\sum \widetilde{Q}_{\alpha\beta}^3\nonumber\right.\\
&&{}-{\textstyle\frac{1}{8}}y_1\sum\widetilde{Q}_{\alpha\beta}^4
-{\textstyle\frac{1}{8}}y_2\sum\widetilde{Q}_{\alpha\beta}^2\widetilde{Q}_{\alpha\gamma}^2\non\\
&&{}-{\textstyle\frac{1}{8}}y_3\sum\widetilde{Q}_{\alpha\beta}\widetilde{Q}_{\beta\gamma}\widetilde{Q}_{\gamma\alpha}^2
-{\textstyle\frac{1}{8}}y_4\sum\widetilde{Q}_{\alpha\beta}^2\widetilde{Q}_{\gamma\delta}^2\non\\
&&\left.{}
-{\textstyle\frac{1}{8}}y_5 \sum \widetilde{Q}_{\alpha\beta}\widetilde{Q}_{\beta\gamma}\widetilde{Q}_{\gamma\delta}
\widetilde{Q}_{\delta\alpha}
\right].
\label{FBR}
\eea
(Here the summations are taken freely over all of the distinct indices displayed in each term.)
Here we included all possible terms of quartic order, as each is generated by the RG; note that $y_1$ has replaced the previous $y$. 
We aim to show that in low dimensions the RG flows take the couplings into a regime where the nature of the transition changes
 qualitatively.

%%%%%%%%%%%%%%%%%%%%%%%%%%%%%%%%%%%%%%%%%%%%%%%
\subsection{Outline and results}

The free energy in eq.\ (\ref{FBR}), evaluated with  $\widetilde{Q}_{\alpha\beta}(\bx)$ independent of $\bx$, gives a Landau theory 
similar to that of GKS, except for the replicon constraint that is now in force, and for the terms with coefficients $y_2$, \ldots, $y_5$.
 Because of 
these differences from the GKS case, our first task is to solve this Landau theory in the various regimes for $y_i$ ($i=1$, \ldots, $5$)
 and for 
$w_1$, $w_2$; this is carried out in Sec.\ \ref{landau} below. Because of the replicon constraint, the Parisi ansatz for RSB applied to 
$\widetilde{Q}_{\alpha\beta}$ leads to a function $\widetilde{q}(x)$ in 
place of $q(x)$, which obeys $\int_0^1 dx\, \widetilde{q}(x) =0$ in place of $q(x)\geq 0$. This change makes little difference in
 practice, and 
the results are very similar to those of GKS summarized above.  The task of including the quartic terms is aided by a paper by Goldbart
 and 
Elderfield \cite{ge}, who considered the full set of quartic terms in the same context as GKS. Similar to what they found, 
the relevant criteria for the extremum to be $1$-RSB (in place of $y\leq 0$ or $y<0$) for $\rt<\rt_c$ are that $y_1-y_3x+y_5x^2\leq
 0$ when 
evaluated at $x=\rho$ for $0<\rho<1$, and  $y_1-y_3+y_5< 0$ when $\rho\geq 1$. It will be convenient to write these criteria simply
 as 
$\widetilde{y}\leq 0$ or $\widetilde{y}<0$ respectively, where $\widetilde{y}=y_1-y_3x_\rho+y_5x_\rho^2$, evaluated at
 $x_\rho=\min(\rho,1)$.
For $\rho>1$, we also find another transition at $\rt=\rt_{\rm d}$ similar to Kirkpatrick and Thirumalai \cite{kirk}, with 
$\rt_{\rm d}>\rt_c$. (They argued that 
this is connected with a dynamical transition.)

In Sec.\ \ref{rgcalc}, we then evaluate the RG flows for our theory at one-loop order in perturbation theory, reproducing the results 
of BR, and extending them 
to include the important part of the flow equations for $y_i$. 

\begin{figure}
\centering
\includegraphics[width=0.8\columnwidth]{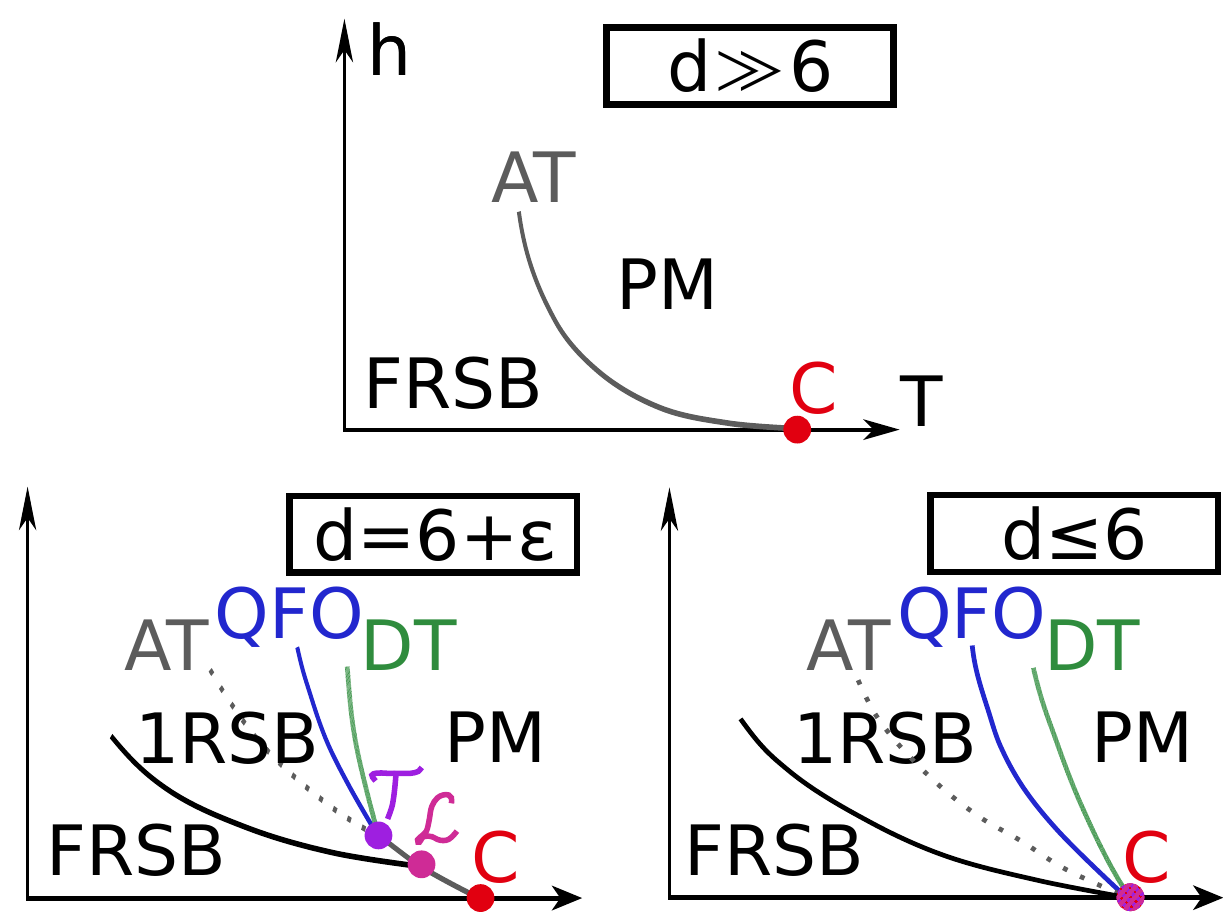}
\caption{(Color online) Schematic phase diagrams for a spin glass in a weak magnetic field near the zero-field critical point $C$, in 
various dimensions $d$ as indicated. 
PM is the paramagnetic phase, FRSB is a full RSB phase, QFO is the quasi-first-order transition, DT is the dynamical transition, 
${\cal L}$ is the Lifshitz-type point, and 
$\cal T$ is the tricritical point. In the places where the AT line appears dotted, it is to indicate where that line would be, though in fact 
it no longer has significance. 
The final panel for $d\leq 6$ is more speculative than the others.}
\label{phdi}
\end{figure}

In Sec.\ \ref{rgan}, we consider the consequences of the flows; the results are summarized in the phase diagrams in Fig.\ \ref{phdi}.
 We find that for $d>8$, 
the transition in weak nonzero magnetic fields takes the same form as 
in mean-field theory: it is continuous and $q(x)$ below the transition is a continuous function. This occurs because all the couplings
 included flow to zero, and, 
making use of 
their initial values (after the crossover from the unreduced or zero-field theory), they do so without reaching either $\rho>1$ or 
$\widetilde{y}<0$. 
For $d\leq 8$, the effective values of the couplings $y_i$ become asymptotic at long length scales to expressions quartic in $w_1$,
 $w_2$, similar to the discussion in FS. 
For $6< d\leq 8$ and at sufficiently small $h$, the $w_i$ couplings flow towards zero, $\rho$ remains small, and $\widetilde{y}$ is again
 positive. 
The behavior is again as in mean-field theory, except for modifications similar to those of FS. 

For $d$ just above $6$, we build on the analysis of MR \cite{mr}. When the magnetic field is not so weak, $\rho$ is driven to larger
 values, 
and there is a Lifshitz-type point $\cal L$ beyond which the phase below the transition is $1$-RSB, while the transition remains
 continuous. At higher field, $\rho$ becomes 
larger than $1$, and the transition to $1$-RSB becomes quasi-first-order. This implies that there is a tricritical point $\cal T$ on the AT
 line at 
$(T_{\cT},h_{\cT})$ (with $h_{\cT}>0$). The tricritical point $\cT$ preempts the multicritical point $M$ in MR, as it occurs at slightly
 lower field in the limit. Likewise, the 
quasi-first-order transition preempts the AT line for $h$ above $h_\cT$. $h_\cT$ tends to zero rapidly as $d\to6^+$. The results
 mentioned are valid within 
perturbation theory up to $h$ larger than $h_\cT$ but of similar order. 

For $d\leq6$, unfortunately we cannot quantitatively analyze the flows in perturbation theory with the present methods. We argue by
 continuity of the phase diagram,
and because no other phase transitions are found within a perturbative analysis, that it is possible that 
the quasi-first-order behavior found for $h>h_\cT$ persists to $d\leq 6$ for all $h>0$. This implies that in these low dimensions, the AT
 line is preempted by 
a quasi-first-order transition, with no divergence of the correlation length or SG susceptibility at the transition. This is consistent with 
a number of simulations, and with high temperature series. Of course, while our arguments may hold for $d$ not too far below $6$, 
we cannot rule out further changes in behavior at lower $d$. We also argue that one or more further transitions, probably including one 
to 
full RSB [i.e.\ with continuous $\qt(x)$],  occur as temperature is lowered further below the transition, as shown in Fig.\ \ref{phdi}. 

In Sec.\ \ref{discuss}, we consider some implications of the results and scenario for the metastate \cite{ns96b,ns_rev,read14} in low 
dimensions
when $1$-RSB occurs below the transition. 
In addition, we further discuss the second solution with $1$-RSB that persists 
to {\em higher} temperatures when the quasi-first-order transition occurs; it is similar to a phase that was discussed before for Potts and 
$p$-spin interaction 
spin glasses, and connected with a dynamical transition \cite{kirk,kw}. We speculate that a dynamical transition may occur in the present 
situation as well. 
Thus we connect the transition in a magnetic field in Ising spin glasses in low dimensions with the REM-like discontinuous transition 
behavior
(with a dynamical transition at higher temperature)
that is now believed to be somewhat generic, and which includes the random first-order transition (RFOT) theory of 
structural glasses \cite{ktw} as well (for a recent review, see Ref.\ \cite{bbccfz}). We discuss the issue of the number of pure states 
that arises in 
these phases.

Sec.\ \ref{concl} is the conclusion, and an Appendix relates quantifying the number of pure states visible in a finite region to mutual 
information. 

We want to point out that our results also apply in cases other than Ising SGs with $2$-spin interactions and a uniform magnetic field. 
For Ising SGs 
in a random magnetic field of mean zero and standard deviation $h$, the same (extended) BR action can be derived. In addition, for a 
SG in which 
the spins are $m$-component unit vectors, inclusion of a mean zero, isotropically-distributed vector-valued random field of standard 
deviation $h$ 
also produces an AT line within mean-field theory \cite{sy}. The same BR action applies there, so that similar results are predicted for 
low-dimensional XY 
and Heisenberg SGs in a random magnetic field. Similarly, we would find the same for a Potts spin glass in a random magnetic field, 
though it is possible 
that there the transition is quasi-first-order even in mean-field theory (depending on $p$). Another family of models are those with $p$-
spin interaction 
among Ising spins \cite{derr}. A particular $3$-spin interaction model has  been mapped to the BR theory \cite{md}, which is reasonable 
in view of the 
lack of inversion symmetry in the $p$-spin models for $p$ odd  (however our conclusions differ from that work); see Ref.\ \cite{cpr} for 
another approach. 

We also mention that Goldschmidt \cite{goldsch} predicted a fluctuation-driven first-order transition in $p>2$ Potts SGs in zero magnetic 
field for $d<6$. 
The prediction was based on the RG flows of the theory, which run to $\rho>1$; his work predates GKS and does not seem to be a 
complete analysis.

%%%%%%%%%%%%%%%%%%%%%%%%%%%%%%%%%%%%
\section{Landau theory of extended BR action}
\label{landau}

We begin by extremizing the action (\ref{FBR}) with respect to $\bx$-independent $\widetilde{Q}_{\alpha\beta}$, using the Parisi
ansatz \cite{par79}; in other words, we consider Landau theory. We drop a factor of volume, and include a Lagrange multiplier $
\lambda$ for the constraints 
$\sum_\beta\widetilde{Q}_{\alpha\beta}=0$ for each $\alpha$ (it turns out that the same value $\lambda$ is found for each constraint, 
so we do not
include a separate multiplier for each), and divide by $n$ \cite{ea}. Then we need to extremize
\bea
{\cal F}[\{\widetilde{Q}_{\alpha\beta}\},\lambda]
&=& \lim_{n\to0} \frac{1}{n} \left[-{\textstyle\frac{1}{2}}\lambda\sum\widetilde{Q}_{\alpha\beta}-{\textstyle\frac{1}{2}}\tau
\sum \widetilde{Q}_{\alpha\beta}^2 \right.\non \\
&&\left.{}-{\textstyle\frac{1}{6}}w_1
\sum \widetilde{Q}_{\alpha\beta}\widetilde{Q}_{\beta\gamma}\widetilde{Q}_{\gamma\alpha}
-{\textstyle\frac{1}{6}}w_2\sum \widetilde{Q}_{\alpha\beta}^3\nonumber\right.\\
&&{}-{\textstyle\frac{1}{8}}y_1\sum\widetilde{Q}_{\alpha\beta}^4
-{\textstyle\frac{1}{8}}y_2\sum\widetilde{Q}_{\alpha\beta}^2\widetilde{Q}_{\alpha\gamma}^2\non\\
&&{}-{\textstyle\frac{1}{8}}y_3\sum\widetilde{Q}_{\alpha\beta}\widetilde{Q}_{\beta\gamma}\widetilde{Q}_{\gamma\alpha}^2
-{\textstyle\frac{1}{8}}y_4\sum\widetilde{Q}_{\alpha\beta}^2\widetilde{Q}_{\gamma\delta}^2\non\\
&&\left.{}
-{\textstyle\frac{1}{8}}y_5 \sum \widetilde{Q}_{\alpha\beta}\widetilde{Q}_{\beta\gamma}
\widetilde{Q}_{\gamma\delta}\widetilde{Q}_{\delta\alpha}
\right]
\label{Flandau}
\eea
with respect to $\lambda$ and $\widetilde{Q}_{\alpha\beta}$, where $\widetilde{Q}_{\alpha\alpha}=0$ and 
$\widetilde{Q}_{\alpha\beta}=\widetilde{Q}_{\beta\alpha}$; we set $\tau=-\widetilde{r}/2$ in this section to simplify writing;
$\tau\propto T_{AT}(h)-T$ is positive for $T<T_{AT}(h)$. 
The value of $\cal F$ at the extremum gives the physical free energy density (into which inverse temperature has been absorbed in both
Landau-Ginzburg and Landau theory).

The Parisi ansatz involves dividing the symmetric matrix $Q_{\alpha\beta}$ into square blocks of equal size, and setting all matrix 
elements in the off-diagonal blocks 
to one value, and those in the diagonal blocks to another value, except for the entries on the diagonal which are zero. This is then 
repeated in the same
way in each diagonal block (replacing the values originally placed there), and iterated so that it is done say $k$ times in total (giving $k$-
step RSB, also 
called $k$-RSB). Formally,
this means that we choose numbers $1=m_0\leq m_1\leq m_2\leq\cdots\leq m_k\leq m_{k+1}=n$, where $m_{i+1}/m_i$ is an integer 
for $i=0$, \ldots, $k$.
Then ${Q}_{\alpha\alpha}=0$ (all $\alpha$), and 
\be
{Q}_{\alpha\beta}=q_i \hbox{ for } \left\lceil \frac{\alpha}{m_i}\right\rceil\neq  \left\lceil \frac{\beta}{m_i}\right\rceil, 
 \left\lceil \frac{\alpha}{m_{i+1}}\right\rceil = \left\lceil \frac{\beta}{m_{i+1}}\right\rceil
\ee
for $\alpha\neq\beta$ and $i=0$, \ldots $k$.
(Here $\lceil a \rceil$ is the ceiling function, the least integer greater than or equal to $a$.) When $n\to0$, the $m_i$s become numbers 
$m_i=x_i$ between $x_{k+1}=0$ and $x_0=1$ and obey the reverse inequalities. Defining a function of these numbers by 
$q(x_i)=q_i$, in the limit $k\to\infty$ 
(and assuming the $x_i$ fill the interval $[0,1]$) we obtain a function $q(x)$. Alternatively, for $k$ finite, we can define $q(x)$ for all 
$x$ to be piecewise 
constant with steps at $x=x_i$, with $q(x)=q_i$ for $x\in [x_{i+1},x_i)$. We note that, as $q(x)$ turns out to be monotonically 
increasing, we can use it 
to define a probability measure on overlaps $q$ (normalized per site) of pure states \cite{par83}; using the function $x\to q(x)$, the 
measure assigned to 
an interval $[q_1,q_2]$ is the Lebesgue measure of $x$ in the inverse image (which is again an interval) of $[q_1,q_2]$.
[Equivalently, $(dq/dx)^{-1}$ (with $\delta$-functions at $q$ for which 
$q(x)$ has zero derivative) can be viewed as the probability density of overlaps $q$ of pure states \cite{par83}.]

In the Parisi ansatz as described, one imposes $q(x)\geq0$. In our case, we assumed (in deriving the BR action \cite{br}) that there is 
a replica symmetric part $Q$, and set $Q_{\alpha\beta}=Q+\widetilde{Q}_{\alpha\beta}$ for $\alpha\neq \beta$, where $\sum_\beta
\widetilde{Q}_{\alpha\beta}=0$ (the replicon constraint). In terms of $q(x)$, we can associate a function $\widetilde{q}(x)$ with 
$\widetilde{Q}_{\alpha\beta}$ exactly as described above for $Q_{\alpha\beta}$, and then
\be
\widetilde{q}(x)=q(x)-Q=q(x)-\int_0^1\! dx\,q(x),
\ee
as the replicon constraint implies that 
\be
\int_0^1 \!dx\,\widetilde{q}(x)=0.
\label{repcons}
\ee 
We will assume $Q$ is nonzero, and that $|\widetilde{q}(x)|$ is 
smaller than $Q$ at all $x$. Then we can ignore the condition $q(x)\geq 0$, but we include the condition $\int_0^1 \!dx\,\widetilde{q}
(x)=0$ instead.

Now evaluating the functional $\cal F$, we find \cite{par79,ge}
\bea
 {\cal F}&=&\half\int_0^1\!dx\,\left[\lambda\qt(x)+\tau\qt(x)^2 +{\textstyle\frac{1}{3}}w_2\qt(x)^3+{\textstyle\frac{1}{4}}
y_1\qt(x)^4
\vphantom{\int}\right.\non\\
&&{}-{\textstyle\frac{1}{3}}w_1\qt(x)\left\{2\langle\qt\rangle\qt(x)+\int_0^x\!dx'\,\left(\qt(x)-\qt(x')\right)^2\right\}\non\\
&&{}+{\textstyle\frac{1}{4}}y_2\left\{\qt(x)^4-2\qt(x)^2\langle\qt^2\rangle\vphantom{\int}\right.\non\\
&&{}-\left.\int_0^x\!dx'\,\left(\qt(x)^2-\qt(x')^2\right)^2\right\}\non\\
&&{}-{\textstyle\frac{1}{4}}y_3\left\{2\qt(x)^3\langle\qt\rangle+\qt(x)^2\int_0^x\!dx'\,\left(\qt(x)-\qt(x')\right)^2\right\}\non\\
&&{}-{\textstyle\frac{1}{4}}y_5\left(\langle\qt^2\rangle^2-\left\{4\qt(x)^2\langle\qt\rangle^2\vphantom{\int}\right.\right.\non\\
&&{}+4\qt(x)\langle\qt\rangle\int_0^x\!dx'\,\left(\qt(x)-\qt(x')\right)^2\non\\
&&{}+\left.\left.\left.\int_0^x\!dx'\,\int_0^x\!dx''\,\left(\qt(x)-\qt(x')\right)^2\right.\right.\right.\non\\
&&{}\qquad\left.\left.\left.\vphantom{\int}\times\left(\qt(x)-\qt(x'')\right)^2\right\}
\right)\right].\non\\
&&
\eea
Here $\langle f\rangle=\int_0^1\!dx\,f(x)$ for a function $f(x)$ on $x\in[0,1]$. The term with coefficient $y_4$ disappears on inserting 
the 
Parisi ansatz, as $\left(\sum_{\alpha\beta}\widetilde{Q}_{\alpha\beta}^2\right)^2$ is of order $n^2$. We note that this term 
represents randomness in the 
mass-squared term $\rt$, and may not be negligible in general, even though it drops out of Landau theory. 

%%%%%%%%%%%%%%%%%%%%%%%%%%%%%%%%%%%%%%%%%
\subsection{Piecewise-differentiable $\qt(x)$}

We can find variational equations for an extremum, assuming that we can apply ordinary functional differentiation. The derivation is 
tedious but 
straightforward, and resembles Refs.\ \cite{par79,ge}. Varying $\lambda$ produces the constraint (\ref{repcons}); this may be used 
freely but 
only {\em after} varying $\cal F$ with respect to $\qt(x)$. Given the variational equation, which is somewhat complicated because of the 
quartic terms in $\cal F$, 
we can obtain simpler equations by applying the operator
\be
\frac{1}{\qt'(x)}\frac{d}{dx},
\ee
as in Refs.\ \cite{par79,ge}, for $x$ at which $\qt'(x)\neq 0$. 
Applying this operator twice, we then find that if $\qt'(x)$ is nonzero at some value of $x$, and if $\qt(x)\to0$ as $\tau$ approaches its 
critical value $\tau_c=0$ from above (i.e.\ for a second-order transition), then we must have
\be
w_2-w_1x={\cal O}(\tau).
\ee
As $\tau\to0$, this can be satisfied only at $x=w_2/w_1\leq 1$ \cite{ge}. Hence such a continuous transition, into a phase with $\qt(x)$ 
differentiable and 
non-constant at $x=w_2/w_1$ just below the transition, can occur only if $0\leq \rho \leq 1$ (where $\rho=w_2/w_1$). If we take an 
additional 
derivative with respect to $x$ 
before letting $\tau\to0$, then we find similarly that as $\tau\to0$, we must have either $\qt'(x)=0$ or \cite{ge}
\be
\qt'(x)\to\frac{w_1/3}{y_1-y_3x+y_5x^2}.
\ee
As $\qt(x)$ must be monotonically increasing in $x$ (for example, to give the interpretation as a probability, mentioned above), and 
because 
both $w_1$ and $w_2$ are positive throughout the paper, we conclude that, defining $\widetilde{y}=y_1-y_3\rho+y_5\rho^2$ for $
\rho\leq1$,
a nonzero finite slope of $\qt(x)$ is possible just below the continuous transition only if $0\leq\rho\leq 1$ and $\widetilde{y}>0$. 
Otherwise, a piece-wise 
constant solution is the only non-trivial possibility for a continuous transition (a constant solution would have to be $\qt(x)=0$, which 
means no 
RSB, and is unstable for $\tau>0$).

Next, we  investigate step-function possibilities when $\widetilde{y}\leq0$. We only consider $\rho>0$. For $\rho<1$, we expect a 
single step to occur, at $x_1=\rho$ as 
$\tau\to\tau_c^+$, in place of the non-zero slope part of $\qt(x)$. A single step means a one-step or ($k=$) $1$-RSB solution.

%%%%%%%%%%%%%%%%%%%%%%%%%%%%%%%%%%%%%
\subsection{$1$-RSB: continuous transition at $\rho<1$}

The preferred way to consider a step-function (or $1$-RSB) solution is to substitute the step-function or $1$-RSB matrix into the general 
Parisi functional $\cal F$. 
The function has the assumed form
\be
\qt(x)=\left\{\begin{array}{ll}q_0,\hbox{ $x<x_1$,}\\q_1, \hbox{ $x\geq x_1$,}\end{array}\right.
\ee
where $x_1\leq 1$. Then the three parameters $q_0$, $q_1$, and $x_1$ (as well as $\lambda$) can be varied to extremize $\cal F$. 
We note that if we used the variational derivative expressions of the previous section, and then substituted the step function, the 
equation
that will be obtained below by varying $x_1$ is not immediately obtained, unless some additional procedure is used. This is why direct 
evaluation of $\cal F$ for
the step function is the simplest procedure.

$\cal F$ can be evaluated for the step function to give
\bea
{\cal F}&=&\half\lambda\langle\qt\rangle+{\textstyle\frac{1}{2}}\tau\langle \qt^2\rangle+{\textstyle\frac{1}{6}}w_2\langle 
\qt^3\rangle
+{\textstyle\frac{1}{8}}y_1\langle \qt^4\rangle\non\\
&&{}-{\textstyle\frac{1}{6}}w_1\left[2\langle\qt\rangle\langle\qt^2\rangle+x_1(1-x_1)q_1(q_1-q_0)^2\right]\non\\
&&{}+{\textstyle\frac{1}{8}}y_2\left[\langle\qt^4\rangle-2\langle\qt^2\rangle^2-x_1(1-x_1)(q_1^2-q_0^2)^2\right]\non\\
&&{}-{\textstyle\frac{1}{8}}y_3\left[2\langle\qt\rangle\langle\qt^3\rangle+x_1(1-x_1)q_1^2(q_1-q_0)^2\right]\non\\
&&{}-{\textstyle\frac{1}{8}}y_5\left[\langle\qt^2\rangle^2-4\langle\qt\rangle^2\langle\qt^2\rangle-4\langle\qt\rangle x_1(1-
x_1)q_1(q_1-q_0)^2\right.\non\\
&&{}-\left.x_1^2(1-x_1)(q_1-q_0)^4\right].
\eea
Other than $\langle\qt\rangle=0$, the equations obtained by varying the parameters in $\cal F$ are somewhat involved. We will 
explicitly solve them in some limiting cases.

First we will suppose that the transition is continuous, and that the quartic terms can be dropped to leading order in $\tau$ (we will see 
that $\tau_c=0$). 
(Note that this is consistent, because $\widetilde{y}=0$ can only lead to a step-function solution.) In addition to $\langle\qt\rangle=0$, 
varying 
$q_0$, $q_1$, and $x_1$ leads (using $x_1$, $1-x_1\neq 0$) to
\bea
0&=&\lambda+2\tau q_0+w_2q_0^2,\\
0&=&\lambda+ 2\tau q_1+w_2 q_1^2-w_1x_1(q_1-q_0)^2,\\
0&=&\lambda(q_0-q_1)+\tau(q_0^2-q_1^2)+{\textstyle\frac{1}{3}}w_2(q_0^3-q_1^3)\\
&&{}-{\textstyle\frac{1}{3}}w_1\left[2(q_0-q_1)\langle\qt^2\rangle+(1-2x_1)q_1(q_1-q_0)^2\right].\non
\eea
Solving, we find to leading order in $\tau>0$ that
\bea
x_1&=&w_2/w_1=\rho,\\
q_0&=&-\frac{\tau}{w_2},\\
q_1&=&\frac{\tau}{w_1-w_2},\\
\lambda&=&\frac{\tau^2}{w_2}.
\eea
These results make sense provided $0<\rho<1$. They are very similar to corresponding ones of GKS \cite{gks}. Higher-order 
terms can be found as power series in $\tau$.

We also mention here that there is a continuous transition for $\rho<1$ and $\tau>0$ from full RSB to $1$-RSB when $\widetilde{y}$ 
changes sign.
We will not discuss this in further detail. The phase boundaries $\tau=0$ (and any $\widetilde{y}$) and $\widetilde{y}=0$ (for $
\tau>0$) for $\rho<1$ 
meet to produce a point analogous to a Lifshitz point (though here the transition for $\tau>0$ is continuous, unlike an ordinary Lifshitz 
point); we refer 
to this as a Lifshitz-type point. 

%%%%%%%%%%%%%%%%%%%%%%%%%%%%%
\subsection{$1$-RSB: discontinuous transition at $\rho>1$ and tricritical case $\rho=1$}
\label{disctr}

In this regime, quartic terms cannot be neglected in general. 
For $\rho>1$, following GKS \cite{gks}, we look for a solution with $x_1\to 1$, $q_1$ tending to a 
positive constant $q_{1c}$, $q_0={\cal O}(1-x_1)$ as $\tau\to\tau_c$ 
(i.e.\ a ``quasi-first-order'' transition, if $\tau_c\neq 0$). Keeping the leading terms in the variational equations, we find $\lambda\sim 
-2\tau_cq_0={\cal O}(1-x_1)$, 
and the system of quadratic equations
\bea
\widetilde{y}q_{1c}^2+(w_2-w_1)q_{1c}+2\tau_c&=&0,\\
{\textstyle\frac{1}{4}}\widetilde{y}q_{1c}^2+{\textstyle\frac{1}{3}}(w_2-w_1)q_{1c}+\tau_c&=&0,
\eea
which has the unique nonzero solution
\bea
\tau_c&=&\frac{1}{9}\frac{(w_2-w_1)^2}{\widetilde{y}},\\
q_{1c}&=&-\frac{6\tau_c}{w_2-w_1}=-\frac{2}{3}\frac{(w_2-w_1)}{\widetilde{y}}.
\eea
Here and below, for $\rho>1$ we define $\widetilde{y}=y_1-y_3+y_5$. Then $\tau_c$ is negative, $q_{1c}$  is positive (and both are 
finite) for $w_2>w_1$, 
$\widetilde{y}<0$. Again, the results are very similar to GKS \cite{gks}. The results are valid within Landau theory provided $w_2-w_1$ 
is sufficiently
small so that $\tau_c$ and $q_{1c}$ are small. We remark that, in spite of the jump in $\qt(x)$ 
at $x=1$ that occurs at $\tau=\tau_c$ (but see the following Sec.\ \ref{dfph}), the latent heat at the transition is zero \cite{gks}. 

For the borderline or tricritical case $\rho=1$, the transition is continuous, but a separate analysis similar to the present section is 
required. 
We find that
\bea
q_0&=&-\frac{2\tau}{w_1},\\
1-x_1&\propto &\frac{(-\widetilde{y}\tau)^{1/2}}{w_1},\\ 
q_1&\propto& \left(-\frac{\tau}{\widetilde{y}}\right)^{1/2},\\
\lambda&=&o(\tau^2)
\eea
as $\tau\to0^+$ for $\widetilde{y}<0$. 

%%%%%%%%%%%%%%%%%%%%%%%%%%%%%
\subsection{Additional step solution at $\rho>1$}
\label{dfph}

A further one-step solution can be found, following Ref.\ \cite{kirk}. As $x_1$ obeys $x_1\leq 1$, we can look for solutions
with $x_1=1$, ignoring the requirement that $\partial {\cal F}/\partial x_1=0$; we still divide the equation 
$\partial{\cal F}/\partial q_1=0$ by $1-x_1$ before solving with $x_1=1$. Then $q_0=\lambda=0$, and a single quadratic is obtained:
\be
\widetilde{y}q_{1}^2+(w_2-w_1)q_{1}+2\tau=0,
\ee
with solution
\be
q_1=\frac{-(w_2-w_1)-\sqrt{(w_2-w_1)^2-8\widetilde{y}\tau}}{2\widetilde{y}}.
\ee
For this to be real and positive when $\widetilde{y}<0$, we require 
\be
\tau>\tau_{\rm d}=\frac{1}{8}\frac{(w_2-w_1)^2}{\widetilde{y}}.
\ee
At the transition at $\tau=\tau_{\rm d}<0$, $q_1$ is nonzero. Thus if $q_1=0$ at $\tau<\tau_{\rm d}$
(i.e.\ high $T$), then $q_1$ becomes nonzero with both a jump and a square-root singularity at $\tau_{\rm d}$. 
$\tau_{\rm d}=\tau_c$ at the tricritical point $\tau_c=0$, $\rho=1$.
Again, these results are valid within Landau theory provided $w_2-w_1$ is sufficiently small, and are similar to 
those in Ref.\ \cite{kirk}. Thermodynamically, this solution is indistinguishable from the paramagnetic or high-temperature one, 
$\qt(x)=0$, as both give ${\cal F}=0$, because $\qt(x)$ differs from $0$ only on a set of measure zero. Within Landau theory, 
we have no way to determine which solution is physical other than by maximizing $\cal F$. Hence it is not clear which of the solutions 
(the paramagnetic one and the present one) is correct in the region $\tau_c>\tau>\tau_{\rm d}$. 

We remark that $0>\tau_c>\tau_{\rm d}$, and that at $\tau=\tau_c$, the values of $q_1$ in the solution here and that 
in Sec.\ \ref{disctr} are the same. At larger $\tau$, the earlier discontinuous $1$-RSB solution has larger $\cal F$, so 
is the physical one. Hence if we accept the $x_1\equiv 1$ solution when $\tau_{\rm d}<\tau<\tau_c$, then at $\tau_c$ there is no jump 
of $q_1$, 
but it is the point such that $x_1$ moves away from $1$ at larger $\tau$ (still with no latent heat). 
We discuss further the meaning of the solution found in this section in Sec.\ \ref{discuss} below. For $\rho=1$, $\tau_c=\tau_{\rm d}
=0$;
this case was discussed at the end of Sec.\ \ref{disctr}.

To avoid confusion, we will refer to the solution found here for $\tau_{\rm d} <\tau<\tau_c$, which has $x_1=1$ throughout, 
as the (dynamically-) frozen phase, reserving the term $1$-RSB for the region $\tau>\tau_c$ (with any value of $\rho$) in which 
$x_1<1$. 

%%%%%%%%%%%%%%%%%%%%%%%%%%%%%%%%%%%%%%%%%
\subsection{$\widetilde{y}\geq 0$ and $\rho>1$, and tetracritical points}

The results so far indicate that most features of the Landau-theory phase diagram can be parametrized using only the three variables
$\tau$, $\rho$, and $\widetilde{y}$. There is also the regime $\rho>1$ and $\widetilde{y}\geq 0$ that we have not discussed. We will 
not investigate this in detail, but only say that, by elimination of other possibilities, and if solutions exist within Landau theory at all, 
then a discontinuous (quasi-first-order) transition should be expected, and 
$\qt(x)$ will be discontinuous (with break-point $x_1\to1$ as $\tau\to\tau_c^+$), but not piecewise constant, thus placing it outside 
the RSB forms we considered here. We might imagine that there would be a transition as $\widetilde{y}$ changes sign when 
$\tau>\tau_c$, and another Lifshitz-type point at $\tau=\tau_c$, this time involving two quasi-first-order boundaries. However, 
as $q_{1c}$ and $-\tau_c\to \infty$ in this limit, Landau theory breaks down before this boundary is reached.

In addition, we mention that the point $\tau=0$, $\rho=1$, $\widetilde{y}=0$ is a tetracritical point, from which all the other phases 
and transitions
mentioned here emerge on changing one or more of these parameters. We will not describe it in detail. Note that we did not consider the 
region 
$\rho\leq 0$ here at all. There is another tetracritical point at $\tau=0$, $\rho=0$, and $\widetilde{y}=0$. 

%%%%%%%%%%%%%%%%%%%%%%%%%%%%%%%%%%%%%
%%%%%%%%%%%%%%%%%%%%%%%%%%%%%%%%%%%%%
\section{Calculation of RG flow equations}
\label{rgcalc}

Next we carry out an RG calculation on the extended BR theory at one-loop order in perturbation theory. The method is a standard one 
\cite{ma}: 
a wavevector cutoff of $1$ is assumed, and Fourier components of fields with wavevectors in a shell just below the cutoff are 
successively
integrated out, followed at each step by rescaling to restore both the cutoff and the coefficient of $(\nabla \widetilde Q)^2$ to $1$. In 
the calculations
reported here, we expand the fluctuations of $\widetilde{Q}_{\alpha\beta}$ around $\widetilde{Q}_{\alpha\beta}=0$. This should be 
valid at least in the 
high-temperature region $\rt>0$ and at the critical point (AT line) $\rt=0$. The calculations
are standard, except for the role of the replicon constraint \cite{br}. The propagator, or zeroth-order two-point correlation function 
$\langle\!\langle  \widetilde{Q}_{\alpha\beta}(\bx)  \widetilde{Q}_{\gamma\delta}({\bf 0})\rangle\!\rangle$,  for 
$\widetilde{Q}_{\alpha\beta}$ 
has Fourier transform 
\be
\frac{S_{\alpha\beta,\gamma\delta}}{\bk^2+\rt}.
\label{rs_prop}
\ee
Here $S$ is a projection operator. In the space of $n\times n$ real matrices with elements ${Q}_{\alpha\beta}$ that are symmetric and 
have zeroes on the diagonal, 
which we equip with norm-square $\sum_{\alpha<\beta} {Q}_{\alpha\beta}^2$, $S$ is the projection onto the subspace
$\sum_\alpha {Q}_{\alpha\beta}=0$. $S$ can be expressed as \cite{br}
\bea
 S_{\alpha\beta,\gamma\delta} &=& 1+\half(\delta_{\alpha\gamma}+\delta_{\alpha\delta}+\delta_{\beta\gamma}+
\delta_{\beta\delta})
                                                        -(\delta_{\alpha\beta}+\delta_{\gamma\delta})\non\\
&&{}-(\delta_{\alpha\beta\gamma}+\delta_{\alpha\beta\delta}
                                                          +\delta_{\alpha\gamma\delta}+\delta_{\beta\gamma\delta})\non\\
                                                       &&{} +(\delta_{\alpha\beta}\delta_{\gamma\delta}
+\delta_{\alpha\gamma}\delta_{\beta\delta}+\delta_{\alpha\delta}\delta_{\beta\gamma}),
\eea
where generalized Kronecker symbols of rank $k$ are defined as $\delta_{\alpha_1\cdots\alpha_k}=1$ if all of $\alpha_i$ ($i=1$, \ldots, 
$k$) 
are equal, and zero otherwise; the limit $n\to0$ has already been taken in the coefficients in this expression. 

Then we obtain the one-loop RG flow equations 
for the effective couplings $\rt(l)$,  $w_i(l)$ ($i=1$, $2$), and $y_i(l)$ ($i=1$, \ldots, $5$)
at length scale $e^l$ (where $l=0$ corresponds to the initial cutoff scale; we often refer to $l$ as a scale,
though it is in fact the logarithm of the length scale), after setting $n=0$:
\bea
\frac{d \tilde{r}}{d l}&=& [2-\tilde{\eta}] \tilde{r}-K_d\frac{(4w_1^2-16w_1w_2+11w_2^2)}{(1+\tilde{r})^2}\non\\
&&{}+\ldots,
\label{rg1}\\
\frac{dw_1}{dl} & = & \half
\left[6-d -3\tilde{\eta}\right]w_1+K_d(14w_1^3-36w_1^2w_2\non\\
&&{}+18w_1w_2^2+w_2^3)+\ldots,
\label{rg2}\\
\frac{dw_2}{dl} & = & \half\left[6-d -3\tilde{\eta}\right]
w_2+K_d(24w_1^2w_2-60w_1w_2^2\non\\
&&{}+34w_2^3)+\ldots,
\label{rg3}\\
\frac{dy_1}{dl}&=&[4-d-2\tilde\eta]y_1+96 K_d w_2^2(w_1-w_2)^2+\ldots,
\label{rg4}\\
\frac{dy_2}{dl}&=&[4-d-2\tilde\eta]y_2+16 K_d(-7w_1^3w_2+15w_1^2w_2^2\non\\
&&{}-11w_1w_2^3+3w_2^4)+\ldots,
\label{rg5}\\
\frac{dy_3}{dl}&=&[4-d-2\tilde\eta]y_3+8 K_d(14w_1^3w_2-28w_1^2w_2^2\non\\
&&{}+13w_1w_2^3+w_2^4)+\ldots,
\label{rg6}\\
\frac{dy_4}{dl}&=&[4-d-2\tilde\eta]y_4+K_d(11w_1^4-32w_1^3w_2\non\\
&&{}+48w_1^2w_2^2-32w_1w_2^3+8w_2^4)+\ldots,
\label{rg7}\\
\frac{dy_5}{dl}&=&[4-d-2\tilde\eta]y_5+K_d(38w_1^4-80w_1^3w_2\non\\
&&{}+40w_1^2w_2^2+w_2^4)+\ldots,
\label{rg8}
\eea
where $\tilde{\eta}=K_d(4w_1^2-16w_1w_2+11w_2^2)(d-4)/d$.
Here the geometric factor $K_d=2/(\Gamma(d/2)(4 \pi)^{d/2})$ arises from integration over
the surface of a sphere in $d$ dimensions in the Fourier integrals. We neglected $\rt$ in denominators arising from one-loop integrals 
after the first equation, eq.\  (\ref{rg1}), and in $\tilde{\eta}$. The $+\ldots$ in the flow equations represent possible further 
one-loop terms, which are of the form $y_i$ ($i=1$, \ldots, $5$) for $\rt$, $w_iy_j$ ($i=1$, $2$, $j=1$, \ldots, $5$) for the cubic 
couplings, and either 
$w_iw_jy_k$ ($i$, $j=1$, $2$, $k=1$, \ldots, $5$) or $y_iy_j$ ($i$, $j=1$, \ldots, $5$)  for the quartic couplings. 
It will be easily seen from the following that these terms are higher order in the weak-coupling regime of interest in this paper, and 
can be dropped (as can the $\tilde\eta$ term in the flow equations for $y_i$ also). We note that the terms quartic in $w_i$ kept in the 
flow equations for $y_i$ are all of the same form as in FS for the unreduced theory. The equations (\ref{rg1}--\ref{rg3}) 
 agree exactly with those of BR \cite{br} (see also Ref.\ \cite{ptd}), though they put $d=6$ in $\tilde\eta$; they (nor, to our 
knowledge, 
any later authors) did not consider the full set of quartic couplings $y_i$. 

%%%%%%%%%%%%%%%%%%%%%%%%%%%%%%%%%%%%%%%
%%%%%%%%%%%%%%%%%%%%%%%%%%%%%%%%%%%%%%%
\section{Analysis of flow equations}
\label{rgan}

In the RG flow equations (\ref{rg1}--\ref{rg8}), all coefficients have been left in their general form, without 
assuming that, say, $d$ is close to $6$. This enables us now 
to analyze the flows in various dimensions. We do this first for the extended BR theory for $d>6$. After that, we apply the results to 
the 
AT line, using the crossover from the unreduced theory to the extended BR theory, beginning with high $d$, and ending with $d<6$.

%%%%%%%%%%%%%%%%%%%%%%%%%%%%%%%%
\subsection{Weak-coupling flows in extended BR theory}

First, we focus on the extended BR Landau-Ginzburg theory. For $d>6$, the Gaussian fixed point, at which coefficients of
all terms beyond the Gaussian ones $(\nabla\widetilde{Q}_{\alpha\beta})^2$ and $\rt(\widetilde{Q}_{\alpha\beta})^2$ are zero,
is stable, that is all those coefficients flow to zero in perturbation theory (at least for sufficiently weak initial values and when only 
finitely many are included).  At the moment we use only the {\em leading} behavior in each coefficient as it flows toward zero. 
First, however, we show that the flows for $y_i$ may take them to the region $\widetilde{y}<0$ if $d\leq8$.

%%%%%%%%%%%%%%%%%%%%%%%%%%%
\subsubsection{Flow of $y_i$}
\label{flowy}

The extended BR Landau-Ginzburg theory is in principle valid anywhere near a transition of AT type (among others), provided 
fluctuations in $\widetilde{Q}_{\alpha\beta}$ are small; hence it may be useful even at large magnetic fields. Perturbation theory
and perturbative RG calculations in $w_i$, $y_i$ and so on are valid if these couplings are, in an appropriate sense, small. For $d>6$, 
there is a nonzero basin of attraction of the Gaussian (zero coupling) fixed point of the RG. Then the theory will be useful
at large length scales if the initial values lie inside this basin of attraction. If $d-6$ is of order one or greater, then this basin contains
all values of the couplings for which perturbation theory would be expected to be valid. For any $d>6$, sufficiently small $w_i$ lie well 
inside 
the basin. For now we take the initial values to be at the scale $l=0$; later this will be replaced with a scale $l_0>0$.

In this regime of linearized flow equations for $w_i$ ($i=1$, $2$), the solutions for $w_i$ are
\be
w_i(l)\approx w_i(0)e^{-\half\eps l},
\ee
(where $\eps=d-6$) so the flow lines are radial in the $w_1$-$w_2$ plane: $\rho(l)=\rho(0)$. For now we will make use only of this 
form. 

For $y_i$ ($i=1$, \ldots, $5$), dropping in the present regime $\tilde{\eta}$ as mentioned already, the equations (\ref{rg4}--\ref{rg8}) 
have the form \cite{fs},
\be
\frac{dy_i}{dl}=[4-d]y_i+A_i(l),
\ee
where, using the preceding approximate solution for $w_i$, each function $A_i(l)$ can be viewed as a known function of $l$ that 
can be read off from the corresponding flow equation. Each equation has general solution
\be
y_i(l)=y_i(0)e^{(4-d)l}
+e^{(4-d)l}\int_0^l A_i(l')e^{(d-4)l'}\,dl'.
\label{yigensol}
\ee
From the flow equations (\ref{rg4}--\ref{rg8}), each $A_i(l)$ is a homogeneous quartic polynomial in $w_1$ and $w_2$, and 
so from the behavior of $w_j$ in the present regime, we then have $A_i(l')\approx A_i(0)e^{-2\eps l'}$. We see that for $d>8$ 
the integral converges as $l\to\infty$, and gives only a correction to the initial value $y_i(0)$; the correction is small if both 
$w_j(0)$ are small. Thus the $y_i$ are proportional to their initial values up to small corrections, and they are simply rescaled 
by the same $l$-dependent factor as they flow towards zero radially in $y_i$-space. (This is in accordance with simple one-loop 
perturbation theory, in which the correction to $y_i$ converges in the infrared for $d>8$, in agreement with Refs.\ \cite{fs,gmb}.) 

For $d\leq8$, the integral does not converge at $l=\infty$, and instead is dominated by its upper limit, to give 
\be
y_i(l)\sim \left\{\begin{array}{ll} A_i(l)/(8-d) &\hbox{ ($d<8$),}\\
                                                  A_i(l)l &\hbox{ ($d=8$),}\end{array}\right.\label{yiasymp}
\ee
as $l\to\infty$ \cite{fs}; in either case, this decays more slowly than the initial value term. Explicitly, these asymptotics are $y_i\sim 
y_i^*$  as $l\to\infty$, 
where, for $d<8$,
\bea
y_1^*&=&96K_dw_1^4\rho^2(\rho-1)^2/(8-d),\\
y_2^*&=&16 K_dw_1^4(-7\rho+15\rho^2-11\rho^3+3\rho^4)/(8-d),\\
y_3^*&=&8 K_dw_1^4(14\rho-28\rho^2+13\rho^3+\rho^4)/(8-d),\\
y_4^*&=&K_dw_1^4(11-32\rho+48\rho^2-32\rho^3\non\\
&&{}+8\rho^4)/(8-d),\\
y_5^*&=&K_dw_1^4(38-80\rho+40\rho^2+\rho^4)/(8-d);
\eea
these are written in terms of $w_1$ and $\rho=w_2/w_1$, and still depend on $l$ in general. The $y_i(0)$ correction is suppressed by a 
factor 
${\cal O}(e^{(d-8)l})$. In the regime considered here of radial flows of $w_i$, the $y_i^*(l)$ describe flow lines that are radial in $y_i$-
space, 
as $\rho$ is effectively constant. As the $y_i^*$s depend only on two parameters, the collection 
of their flow lines form a two-dimensional space; intersected with a sphere of radius $K_dw_1^4$ (i.e.\ setting $w_1^4$ to a constant), 
we obtain 
a curve parametrized by $\rho$. For general initial conditions ($|y_i(0)|<1$) and given $w_j$, all flow lines of $y_i$ asymptotically (at 
large $l$) 
approach the origin along one of these lines. This differs from the case $d>8$, in which the flow lines of $y_i$ do not converge onto a 
single line, 
but approach the origin in $y_i$-space from all directions. For $d=8$, $(8-d)^{-1}$ should be replaced in these expressions by $l$, as in 
eq.\ (\ref{yiasymp}). 
Then the terms in $y_i(l)$ quartic in $w_j$ are larger by $l$ than the $y_i(0)$ terms. This means that the approach 
to $y_i^*(l)$ is logarithmically slow in length scale, but still occurs.

We can now begin to apply the results of Sec.\ \ref{landau}. In all cases we need to evaluate
\be
\widetilde{y}=y_1-y_3x_\rho+y_5x_\rho^2
\ee
where $x_\rho=\min(\rho,1)$. For $d\leq 8$ with $w_i$ small, the $y_i$ tend to their asymptotic behavior $y_i^*(l)$, while $\rho$ does 
not change 
during the flow. When we evaluate $\widetilde{y}(l)$ using these 
$y_i(l)=y_i^*(l)$, we obtain $w_1^4$ times a polynomial in $\rho$, for either $\rho\leq1$ or $\rho\geq 1$. For $ d<8$ and $\rho\leq 
1$, we have 
\be
\widetilde{y}=K_d w_1^4\rho^2 (22-48\rho+32\rho^2-8\rho^3+\rho^4)/(8-d),
\ee
and for $\rho\geq 1$,
\be
\widetilde{y}=K_d w_1^4( 38 - 192\rho + 360\rho^2 - 296\rho^3 + 89\rho^4 )/(8-d).
\ee
Then $\widetilde{y}(l)$ turns out to be positive for small $\rho>0$, but {\em negative} for $0.8418 < \rho < 1.2694$.
{\em This is a key point of the analysis.} The fact that $\widetilde{y}<0$ in this range of $\rho$ values means from the results of 
Sec.\ \ref{landau} that $\qt(x)$ {\em is discontinuous} for $\rt$ below $\rt_c$, giving $1$-RSB. 
This is one of the central results of the paper, and will demonstrate a qualitative change from the behavior in the SK model in dimensions 
near and 
below $d=6$. For $\rho<0.8418$, $\widetilde{y}(l)>0$, and the transition is to full RSB, and is continuous.  For $0.84<\rho<1$ the 
transition 
remains continuous. At $\rho=0.8418$, $\widetilde{y}=0$, producing a Lifshitz-type 
point, with a transition line from full RSB to $1$-RSB extending from it into the region $\rt<0$. As $\widetilde{y}<0$ at $\rho=1$, there 
will be a tricritical 
point there, beyond which the transition becomes discontinuous. At $\rho=1.2694$, again $\widetilde{y}=0$, and we know of no 
solution to Landau theory 
for $\rho\geq 1$, $\widetilde{y}\geq0$.

For $d>8$, the $y_i$ flow essentially radially to zero, and the initial values of $y_i$ determine the behavior, provided $w_j$ are small.

%%%%%%%%%%%%%%%%%%%%%%%%%%%%%%%
\subsubsection{Critical phenomena}
\label{critical}

Next we analyze for $d>6$ the Gaussian fixed point with arbitrary perturbations whose asymptotic flows were discussed in Sec.\ 
\ref{flowy}. 
First, for $\rt$, in eq.\ (\ref{rg1}), we keep only the terms linear in $\rt$ (the zeroth order term is of little interest), and then the AT line 
$T_{AT}(h)$ 
corresponds to $\rt=0$. Then as $\rt(l)$ is the only relevant parameter at the Gaussian fixed point when $d>6$, it grows with $l$. 
We can stop the flows at $l=l^*$ such that the largest mass-square in any propagator is equal to $1$, and then solve the 
Landau-Ginzburg 
theory with the parameters at $l=l^*$ to obtain the phase transition behavior. (The closer the initial $\rt(0)$ is to zero, the larger the 
$l$ 
that will be required.) On the high-temperature side
of the transition, this mass-square will be $\rt>0$, but on the low temperature side its value should be considered further. For $
\rt<\rt_c$, 
strictly in the RG one should use propagators in the expansion of the action about its extremum \cite{ma} which may 
be at nonzero $\widetilde{Q}_{\alpha\beta}(l)$, and the latter should be small. In the regime where the Landau-Ginzburg theory is valid 
(in terms 
of the action at scale $l$), this will always be true sufficiently close to the transition if it is continuous (in fact, when $\rho<1$), and one 
would
hope also if it is weakly first order, meaning that the extremum value of $\widetilde{Q}_{\alpha\beta}$ is small as $\rt\to\rt_c^-$. In 
these cases, 
the only leading order effect of nonzero $\widetilde{Q}_{\alpha\beta}$ will be to make the mass-squared term in the propagator 
positive; it can 
be neglected in the flows themselves. In the cases of the phases of interest in this paper, the $1$-RSB form holds for $\rt<\rt_c$, and 
we have 
checked that the mass-squares are positive (in agreement with GKS \cite{gks}), and that the largest is of order $|\rt|$ as $\rt\to\rt_c^-$ 
in 
most cases; see also Refs.\ \cite{ck,cwilich,hr}. Then in most cases, for the low $T$ side the only change is that we must stop the flows 
at $-\rt=1$ 
(within a numerical factor), and the analysis is essentially the same as for $\rt>0$. Then, once the flow has stopped, we can apply the 
Landau theory 
analysis of Sec.\ \ref{landau}, using the effective values of parameters at scale $l$. We will follow this standard approach. 

For these reasons, we now consider the solution of Landau theory as in Sec.\ \ref{landau}, but using the effective ($l$-dependent) 
values 
of the parameters for $d\leq 8$. (For $d>8$, the critical behavior agrees with Landau theory.) Thus in terms of the effective phase 
diagram of
Landau theory which involved three parameters ($\rt$, $\rho$, and $\widetilde{y}$), the system flows onto a two dimensional surface 
within 
that space, as $\widetilde{y}$ becomes a function of $\rho$. We recall that $\rt=-2\tau$, so here 
$\rt(l)=-2\tau(l)$. In addition to the exposition of Sec.\ \ref{landau} we point out that
if $q$ stands for any of  $\widetilde{Q}_{\alpha\beta}$, $\qt(x)$, $q_0$, or $q_1$, then at zeroth order in perturbation theory we 
have 
\be
q(l)=e^{\half(d-2)l}q(0).
\ee
We will mainly assume that $\widetilde{y}<0$.

For $0<\rho<0.8418$, the transition is to full RSB, and the scaling for $6<d\leq8$ is very similar to that of FS for the zero field case, 
so there is little that we can add here.
For $0.8418<\rho<1$, on inserting $\tau(l^*)=1$ and $w_i(l^*)$ into the expressions in Sec.\ \ref{landau}, we obtain for $\rt<0$ that 
$q_0(l^*)$, 
$q_1(l^*)\sim 1/w(l^*)\sim e^{\half\eps l^*}$ (we neglect numerical coefficients and initial values in this section, and $w$ stands for 
$w_1$, $w_2$, or if $\rho\neq 1$, $|w_1-w_2|$). In terms of $q$ at $l=0$, we find $q_i(0)\sim e^{-2l^*}=\tau(0)$, which we can 
characterize 
with the exponent $\beta=1$. We note that $y_i$ do not appear in the expressions for $q_0$, $q_1$, or $x_1$ at leading order
in $\rt$, so the $y_i$ are irrelevant and not dangerous at this continuous transition (i.e.\ at the Gaussian fixed point for $\rho<1$), 
provided $\widetilde{y}\leq 0$, though they are dangerous in the RSB region as in FS, similar to Refs.\  \cite{ck,cwilich}. ($w_i$ are 
still dangerously irrelevant for $d>6$, and responsible for the breakdown of the hyperscaling relation implied by $\beta=1$.) 

For the cases of the discontinuous (quasi-first-order) or the dynamical transition when $1<\rho<1.2694$, we can sit at $\tau_c$ or 
$\tau_{\rm d}$, 
which are both negative and scale the same way, so the results are the same for both; we consider $\tau_c$. In order to set 
$\tau_c(l^*)\sim (w_2-w_1)^2/\widetilde{y}=-1$, we must have $w_2-w_1\sim e^{-\eps l^*}$, so $\rho-1\sim e^{-\half\eps l^*}$. 
Then $q_{1c}(l^*)\sim  e^{\eps l^*}$, so $q_{1c}(0)\sim \rt_c(0)^{(10-d)/4}$ for $d<8$, where the exponent interpolates the 
values $1/2$ at $d=8$ (as in Landau theory) and $1$ at $d=6$ (as required by hyperscaling). This is an instance of the scaling forms 
found by FS \cite{fs}. 

In all the preceding cases, the largest mass-square in any propagator just below or at the transition is of order $|\rt|$. For the tricritical 
point 
$\rho=1$, where $\rt_c=0$, this is not the case: the largest mass-square on the low-temperature side is instead of order 
$w_1(\rt/\widetilde{y})^{1/2}$ \cite{hr}. This implies that, within fluctuations around mean-field theory, without use of RG, the 
critical exponent for the leading correlation length is $\nu'=1/4$ on the low temperature side for this case, as compared with $\nu=1/2$
on the high-temperature side. (There is also a mass-square that goes to zero in a similar way at the dynamical transition \cite{ck,cwilich}, 
implying a corresponding correlation length exponent $\nu'=1/4$, however there it is subleading to others $\propto \rt_{\rm d}>0$ 
as $\rt\to\rt_{\rm d}$.) We can carry out a similar analysis as before, but stopping the RG flow when $w_1(l^*)[\rt(l^*)/
\widetilde{y}(l^*)]^{1/2}=1$. 
Then $\tau(l^*)\sim e^{-\eps l^*}$, so $\tau(0)\sim e^{-(d-4)l^*}$, and $q_1(0)\sim e^{-2l^*}\sim \tau(0)^{2/(d-4)}$ for $6<d\leq 
8$, 
while $q_0(0)\sim \tau(0)$ for all $d>6$. The largest mass-square (in units for $l=0$) is then $\sim e^{-2l^*}\sim \tau(0)^{2/(d-4)}$. 
Again, 
for $6<d\leq 8$, the exponents for $q_0$, $q_1$ interpolate between their values as $d\to 6^+$ (obeying hyperscaling) and at $d=8$ 
(as in Landau theory), while 
now $\nu'=1/(d-4)$ for $6<d\leq 8$, which interpolates between $\nu'=1/4$ for $d=8$ and $\nu'=1/2$ (in agreement with the scaling 
law $\nu'=\nu$) 
for $d\to 6^+$. These results are analogous to some of those of FS, 
though the precise forms differ, as this particular dependence of the mass-square on $\widetilde{y}$ did not occur in FS, and they did 
not have a $\nu'=1/4$. 

Further information about scaling properties can be obtained as well, analogously to Refs.\ \cite{ck,cwilich}. Note that in some cases 
\cite{ck,cwilich,hr} 
there remain modes in the ordered phase that are still massless at $l=l^*$ (i.e.\ the mass-squares are much smaller than the leading 
ones, and tend to zero at 
$\rt_c=0$ and at $\rt_{\rm d}$). It would be possible, and it is necessary, to restrict to a theory of these modes alone (analogous to 
what BR 
did in constructing their action), and then carry out the RG at $l>l^*$ on this smaller set of modes. While it is possible that this will 
change the scaling of
$q_i(0)$, we do not expect that it does, and we will not consider it further here. 

We have some concerns about the preceding analysis of the scaling behavior at the tricritical point and the discontinuous transition.
In addition to $y_i$, we can similarly treat terms of higher order $k>4$, for example $z_k\sum_{\alpha\beta}
\widetilde{Q}_{\alpha\beta}^k$; for each of these
there are one-loop diagrams that contain a polygon with $k$ sides (for $k=4$, this becomes the ``box'' diagram form as in FS), and for 
$d<2k$ the resulting flows 
approach $z_k(l)\sim \sum_{a=0}^k b_{ka} w_1^{k-a}(l)w_2^a(l)\sim e^{-\half k\eps l}w(0)^k$. (We will not need the detailed form of 
these 
asymptotics as we did for $y_i$.) If we add one of these terms to the action as a perturbation, and look at its effect on $q(l)$, we find a 
negligible effect
of relative order $w(l^*)\sim e^{-\half\eps l^*}$ (for all $k$) in the case of the continuous transition. But for the discontinuous 
transition (as $\rho\to1^+$), 
we estimate a change in $q_{1c}(l^*)$ of relative size $e^{\half (k-4)\eps l^*}$, which increases with $l^*$ for $6<d\leq 8$. Similarly 
for the tricritical point, we find
that the change is of relative order $1$, intermediate between the preceding cases. These results mean that, especially for the 
discontinuous and dynamical transitions,
and possibly also for the tricritical point, these higher-order terms should not be neglected, and it would be better to include all of them 
by summing them up 
to obtain the one-loop fluctuation correction to the free energy. We suspect that the result will be modifications of the scaling for 
$6<d\leq 8$ in at least 
some of these cases, but we will not pursue this here. We emphasize that we do not expect such effects to modify the qualitative form 
of the transitions; 
in particular, the tricritical point will still exist.

%%%%%%%%%%%%%%%%%%%%%%%%%%%%%%%%%%%%%%%%%%%%%%%%%%%%%%%
\subsection{Fate of the AT line}

Now we come to the central points: the application of the preceding results to the AT line. We work downwards in dimension $d$, from 
$d\gg 6$ to $d<6$.

\begin{figure}
\centering
\includegraphics[width=\columnwidth]{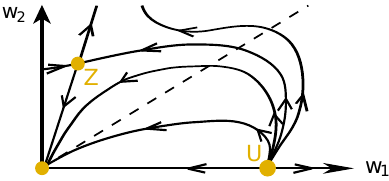}
\caption{(Color online) RG flows for $w_1$, $w_2$ for $d=6+\eps$, $\eps>0$. The flows shown are correct topologically, but not 
drawn to scale. 
$U$ and $Z$ are fixed points; $Z$ is on the fixed line $w_2/w_1=\rho^*=14.379$. $w_2=w_1$ ($\rho=1$) is shown as the dashed 
line. 
The unique flow line that approaches $\rho=1$ tangentially as $l\to\infty$ is shown. The flow line asymptotically tangent to $\rho 
=0.8418$ 
is qualitatively similar to that one.}
\label{fldi}
\end{figure}

First, we describe the effects of the nonlinear terms in the flows for $w_i$. The flows for $d>6$ were shown explicitly in Ref.\ \cite{mb} 
(see Fig.\ \ref{fldi}). There 
is a weak coupling region inside which flows go to the origin, separated by a lobe-shaped separatrix from the region where flows go to 
infinity (within the 
leading one-loop RG). There are two pairs 
of fixed points on the separatrix (the two members of each pair are related by a symmetry, $w_i\to-w_i$, $i=1$, $2$). One pair of fixed 
points are at $U$, $-U$ 
defined by $w_1=\pm \sqrt{\eps/(24K_d)}$, $w_2=0$, and both are unstable under a perturbation that remains on the separatrix. The 
other pair $Z$, $-Z$ 
are on the fixed line $\rho=\rho^*=14.379$, and both are attractive on the separatrix. When the nonlinearities are important within 
perturbation theory, 
that is when $d\leq 6+{\cal O}(\eps)$ and $w_i$ is order $\eps^{1/2}$ or greater, the value of $\rho$ 
changes during the flow (unless $\rho= 0$, which is also a fixed line). In the strong-coupling region, or for all nonzero couplings if 
$d<6$, $\rho$ flows towards
the fixed value, $\rho^*$. Moreover, we should reconsider the RG flows of $y_i$ in light 
of the flows of $w_i$. We can no longer carry out the integration of the flow equations for $y_i$ directly. But, as each $y_i$ has 
dimension $4-d$ to zeroth order, while 
$w_i$ has dimension $(6-d)/2=-\eps/2$ (and the latter is still of the correct order in the vicinity of the separatrix), the integral in the 
solution 
for $y_i(l)$, eq.\ (\ref{yigensol}), contains a kernel that decays rapidly (as $l'$ {\em decreases}) compared with the inverse rate of 
change of $A_i(l)$. 
This means that for large $l$ and 
with $\eps$ small, again $y_i\sim y_i^*$ as $l\to\infty$. Hence we know the expressions for $y_i^*$ in terms of $w_1$ and $\rho$, and 
the resulting $\widetilde{y}(l)$
is negative for $0.8418 < \rho < 1.2694$, but we will need to consider further the flows of $w_j$ (or of $w_1$ and $\rho$).

%%%%%%%%%%%%%%%%%%%%%%%%%%%%%%
\subsubsection{$d>6$ at sufficiently weak magnetic field}

In order to describe the AT line, we will need some results about the unreduced Landau-Ginzburg theory as in Ref.\ \cite{hlc}, and the 
crossover to the
BR reduced theory. For the unreduced theory, the field is $Q_{\alpha\beta}$ without the restriction to the replicon subspace, and the 
action is
\bea
F_{\rm un}[\{{Q}_{\alpha\beta}\}]
&=& \int   d^dx\,   \left[{\textstyle\frac{1}{4}}\sum(\nabla
{Q}_{\alpha\beta})^2-{\textstyle\frac{1}{2}}h^2\sum{Q}_{\alpha\beta}
 \right.  \non \\
&&\left.{}+{\textstyle\frac{1}{4}}{r}
\sum {Q}_{\alpha\beta}^2-{\textstyle\frac{1}{6}}w
\sum{Q}_{\alpha\beta}{Q}_{\beta\gamma}{Q}_{\gamma\alpha}
\nonumber\right.\\
&&{}-{\textstyle\frac{1}{8}}y\sum{Q}_{\alpha\beta}^4
+{\textstyle\frac{1}{4}}x\sum{Q}_{\alpha\beta}^2{Q}_{\alpha\gamma}^2\non\\
&&\left.{}
-{\textstyle\frac{1}{8}}u \sum{Q}_{\alpha\beta}{Q}_{\beta\gamma}{Q}_{\gamma\delta}{Q}_{\delta\alpha}
\right].
\label{Fun}
\eea
For $d>6$ and zero magnetic field, all couplings (of order higher than quadratic)
flow towards zero at leading order in perturbation theory (including the nonlinear terms for $dw/dl$). When $h$ is nonzero, at some 
scale $l=l_0$, 
$|r(l_0)|=1$ becomes of order $1$, and the nonreplicon modes become massive. That is when the crossover to the BR reduced theory 
occurs. 
The initial values of the couplings in the BR action, at the initial scale that is now $l_0$, are then determined as follows \cite{br}, with all 
parameters taking 
their effective values at $l=l_0$. In non-zero magnetic field 
there is a replica symmetric order parameter $Q_{\alpha\beta}=Q$ for $\alpha\neq\beta$ \cite{sk}, and BR showed that the result of 
expanding 
the action in terms of $\widetilde{Q}_{\alpha\beta}=Q_{\alpha\beta}-Q$ (for $\alpha\neq \beta$) and then imposing the replicon 
constraint on 
$\widetilde{Q}_{\alpha\beta}$ is to produce the values
\bea
\rt&=&r+2wQ,\\
w_1&=&w-3uQ,\\
w_2&=&3yQ,
\eea
and further $y_1$, $y_2$, and $y_5$ will be the same as their counterparts $y$, $-2x$, and $u$ in the unreduced 
action $F_{\rm un}$ above [this $x$ has nothing to do with Parisi's $x$ in $q(x)$], up to terms of higher order in $Q$ (that involve 
terms higher than quartic order in the unreduced theory); $Q=-r/(2w)=[h^2/(2y)]^{1/3}$ 
on the AT line \cite{br}. In addition, the initial values of $y$, $x$, and $u$ at $l=0$ (not $l=l_0$!) are positive.  $y_3=0$  in the 
unreduced theory 
because of symmetry at $h=0$, so for us it is of higher order, while we have seen that $y_4$ is not needed in Landau theory.

For the location of the AT line for $d>6$, we have $h(l_0)^2=-y(l_0)r(l_0)^3/(4w(l_0)^3)\sim e^{\half(d-10)l_0}$ for $d\geq8$, and 
$\sim e^{-\half(d-6)l_0}$ for $6<d\leq 8$. At weak coupling, $h(l)^2\sim e^{\half(d+2)l}h(0)^2$. Then in terms of $h(0)$ and $-r(0)
\sim e^{-2l_0}$, 
this gives $h(0)^2\sim -r(0)^3$ for $d\geq 8$, and $h(0)^2\sim [-r(0)]^{(d-2)/2}$ for $6\leq d\leq 8$ as $h(0)\to0$, in agreement with 
FS and Green {\it et al.} 
\cite{fs,gmb}. Note that for now we consider the limit $h(0)\to0$ ($l_0\to\infty$) at fixed $d$. 

For the initial values of couplings in the BR theory, we use $Q(l_0)\sim w(l_0)^{-1}\sim e^{\half\eps l_0}$.  Then for $d>6$, we find 
$w_1(l_0)\sim 
w(l_0)\sim e^{-\half\eps l_0}w(0)$, 
and, for $d\geq 8$, for the ratio $\rho(l_0)\sim e^{-2l_0}\sim -r(0)$. For $6<d\leq 8$, we have instead $\rho(l_0)\sim e^{-\eps l_0}\sim 
[-r(0)]^{\eps/2}$, 
or $w_2(l_0)\sim w_1(l_0)^3$. Taking into account the relation of $r(0)$ and $h(0)$ given by the AT line, these mean that $w_1(l_0)$ 
and $\rho(l_0)$ 
are small at weak magnetic field on the AT line. Note that $r(0)\propto T_{AT}(h)-T_{AT}(0)<0$ for positive $h$.

We can now draw conclusions about the transitions and ordered phase below the AT line for $d>6$ at weak magnetic field. 
For any $d>6$, when the crossover to the extended BR theory occurs, initial values (now at $l=l_0$) of $w_1$ and $\rho$ are small, as 
are the values of $y_i(l_0)$. 
For $d>8$, as $\rho$ is small and does not flow at weak coupling, and as the ratios of $y_i$ do not flow either, for $\widetilde{y}$ the 
most important term 
is $y_1$, so $\widetilde{y}\approx y>0$; it remains positive under the flow at larger $l$. Hence at $l^*$, when the Landau theory can 
be used at $\rt<0$, 
$\widetilde{y}(l^*)>0$, and the transition, which occurs at $\rt_c=0$, is continuous; the Parisi $q(x)$ function [or $\qt(x)$] is 
continuous below the transition. 
Thus, unsurprisingly, the behavior for $d>8$ at small $h$ is that conventionally expected on the AT line.

For $6<d\leq 8$ at sufficiently weak magnetic field, as we have seen the flows take $y_i$ to $y_i^*(l)$, and $\rho$ is invariant during 
the flow.
As $\rho$ is initially small and positive, it remains so, and we are in the regime in which $\widetilde{y}>0$. Even though the flow could 
have driven 
an initial positive $\widetilde{y}$ to negative values, for the initial values in question this does not happen. Hence for $d>6$ and at 
sufficiently weak magnetic 
field, the transition remains a continuous one to full RSB, as for $d>8$. The transition line is still the AT line, 
$T_c(h)=T_{AT}(h)$, with the  FS scaling properties at weak magnetic field, as mentioned just now.

%%%%%%%%%%%%%%%%%%%%%%%%%%%%%%%%%%%%
\subsubsection{Tricritical and Lifshitz-type point on AT line for $d\to6^+$}

For dimensions close to $6$, we consider here $\eps>0$ and small. The flows of $w_1$, $w_2$ in this region were 
considered by BR and others \cite{br,mb,mr}. When these couplings are of order $\varepsilon^{1/2}$ or greater, 
the flows are no longer given by the exponential decay that we saw for larger $d$ (or at weak fields), but instead 
the flow lines exhibit curvature (see Fig.\ \ref{fldi}). The crossover and initial values at $l_0$ were discussed by MR \cite{mr}. From 
those results,
we must now consider the limit as $\eps\to0$ with $\eps l_0$ fixed. We find also from MR that the value of the ratio initially is 
$\rho(l_0)={\cal O}(\eps)$ \cite{mr} (because $w_1(l_0)\sim\eps^{1/2}$). The important region once the BR flows apply is close to the 
unstable fixed point $U$; the initial values lie on a line of slope that tends to zero as $\eps\to0$. 
The RG flows take $\rho>0$ to larger values. For flows inside the separatrix, the flows initially increase 
$\rho$ and possibly $w_1$, $w_2$, but eventually fall back towards the origin along radial lines with constant slope $=\lim_{l\to\infty}
\rho(l)$; 
these were discussed earlier. As $w_1(l_0)$ increases, the value of $\rho(l\to\infty)$ increases and eventually exceeds $1$. Once $
\rho$ 
is larger than $0.8418$, $\widetilde{y}(l\to\infty)$ is negative. This implies that the phase just below the transition becomes $1$-RSB, 
while the transition is still at $\rt_c(l\to\infty)=0$ (the AT line), and is continuous.

There is a unique flow line that leaves $U$ and approaches the origin with slope $\rho=0.8418$ asymptotically as $l\to\infty$. 
Because $\widetilde{y}(l)<0$ for $\rho(l)$ close to $1$, initial values that intersect this flow line define the Lifshitz-type point $\cal L$ on 
the AT line in the $T$--$h$ plane. 
The form of the flows (in particular, the eigenvalues of the linearized equations at $U$) implies that, on going back along the flows (i.e.\ 
as $l\to-\infty$), 
this flow line approaches $U$ asymptotically tangent to the separatrix. As $\rho(l_0)\propto \eps$ is small, we see that as $\eps\to0$, 
the 
difference between $w_1/\eps^{1/2}$ on this flow and on the separatrix is a higher-order correction. Consequently, to leading order, 
we 
find the exact same asymptotic behavior of the location of this point as $\eps\to0$ as for the multicritical point $M$ in MR \cite{mr} ($M$ 
arises 
from the flow into $Z$). There is a similar flow line that approaches $\rho=1$ as $l\to\infty$ (shown in Fig.\ \ref{fldi}). It defines the 
tricritical point $\cT$ at $(T_\cT,h_\cT)$ in the $T$--$h$ plane. It too has the same 
asymptotic behavior as $M$ in MR. Likewise, there is a third flow line that leaves $U$ and approaches the origin with asymptotic slope 
$1.2694$. 
But because this region undergoes a discontinuous transition at $\rt(l^*)=1$, while $\rt_c$ in Landau theory at this transition diverges 
as 
$\widetilde{y}\to0$, the asymptotic region is not reached, and $\rho(l^*)$ never exceeds $1.2694$ for any flow starting at $h>h_\cT$. 
Thus, for both points $\cal L$ and $\cT$, we then have the same leading asymptotic location as for $\cT$ (or $M$ in MR \cite{mr}), that 
is
\bea
r(0)_{\cal T}&=& \eps^{-5/3}\left( e^{-\eps l_{0\cal T}}\right)^{2/\eps}\left[2K_d w_0^2(1-e^{-\eps l_{0\cal T}})\right]^{5/3}\!\!,
\quad\quad\\
h(0)_{\cal T}^2&=&\frac{A\eps^4 r(0)_M^2 w_0 e^{-\eps l_{0\cal T}}}{8\left[2K_d w_0^2(1-e^{-\eps l_{0\cal T}})\right]^4}
\eea
(where $A$ is a constant, and we restored the factors $K_d$ for consistency; $K_d$ can be evaluated at $d=6$ here), as $\eps\to0$ 
with $\eps l_{0\cal T}$ 
held constant at $\eps l_{0\cal T}=\ln 13=2.565\ldots$ \cite{mr}. Thus, like $M$, the tricritical point $\cT$ and the Lifshitz-type point 
$\cal L$ tend to the 
zero magnetic field critical point $C$ as $d\to 6^+$, and do so exponentially fast in $\eps$.

Although the leading asymptotic positions of $\cal T$ and of $M$ are the same, nonetheless $\cal T$ preempts $M$ as $h$ is increased 
on the AT line 
(i.e.\ occurs  at smaller $h$ and $T_c-T$), because the value of $w_1$ at $l_0$ is smaller for the flow to $\cal T$. (From here on, we 
write $h(0)$ as 
simply $h$, and similarly 
$h_\cT(0)$ as $h_\cT$.) At $h$ slightly larger than $h_{\cal T}$, the discontinuous transition sets in,
and occurs at $\rt>0$, that is $T_c(h)$ is above the AT line $T_{AT}(h)$: {\em it preempts the AT transition} in this region. In addition, 
the dynamical 
transition at $T_{\rm d}(h)$ (if it exists) occurs for $h>h_{\cal T}$, and $T_{\rm d}(h)>T_c(h)>T_{AT}(h)$ for $h>h_{\cal T}$. 

In the vicinity of $\cal T$ in the $T$--$h$ plane, the ordered phase below (but not too far below) $T_c(h)$ is again of the $1$-RSB form.
For the critical behavior at and near the tricritical point $\cal T$, at $\eps$ fixed and small, but $T$ or $h$ approaching $\cal T$, we have 
the critical properties,
described above. For $h$ just above $h_{\cal T}$, the asymptotic $\rho(l)-1$ is small, so the scaling for the jump in $q_1$ applies. We 
note that the 
critical exponents at $M$ were non-trivial \cite{mr}, while simple behavior is found for $\cal T$, because it is controlled by the Gaussian 
fixed point. 
The latter behavior seems more natural than having nontrivial exponents {\em above} the upper critical dimension, here $d=6$. 

For larger $h$, the initial values at $l_0$ would hit, then pass outside of, the separatrix. As the value of $q_1(l^*)$ when the flow stops 
becomes of order one, 
we eventually pass out of the domain of validity of Landau theory itself (even if $w_i(l^*)$ are still small), and should not trust the 
results beyond such a point. 
As the flows stop before $\rho=\rho^*$, the fixed point $Z$ that controlled the point $M$ in MR will not be reached (it plays no role in 
the analysis). 
Hence the results may be valid even for initial values somewhat outside the separatrix. But with the present methods the results cannot 
be justified beyond 
a value of $h$ that is of the same order as $h_\cT$ as $\eps\to0$. 

The fate of $\cal T$ and $\cal L$ as $d$ increases to $d-6$ of order one or larger is not known. At some dimension $d>6$ one or both 
may reach the 
zero-temperature line and disappear. It is also possible that one or both of them persists at high field, even above $d=8$. 

%%%%%%%%%%%%%%%%%%%%%%%%%%%%%%%%%%%%%
\subsubsection{$d\leq6$}

We have shown that a discontinuous (quasi-first-order) transition occurs for $d$ slightly above $6$ and at magnetic field $h>h_\cT$. 
Although we do not see a transition to other behavior at sufficiently large $h$, we cannot rule it out. Likewise, because of the continuity 
of RG flows 
as a function of $d$  as well as of other parameters (away from fixed points), it is possible that the discontinuous transition persists to 
$d=6$ and below, 
here for all (sufficiently small) $h>0$, though other behavior might set in for $6-d$ sufficiently large, and strictly speaking the region of 
validity 
of the analysis above shrinks to zero as $d\to6^+$. Again, the discontinuous transition (and the dynamical transition, if it exists) would 
preempt 
the AT line, and would do so now for {\em all} $h>0$.

We now consider the quantitative analysis of the transition at small $h$ for $d\leq 6$. We only consider $d$ not far below $6$. Then the 
flows of $y_i$ for 
$d=6+\eps$ still apply, and $y_i\sim y_i^*$ for $l\to\infty$. 

For $d=6$, the RG flows for $w_i$ exhibit a scaling property: they are invariant if we rescale all of $w_1$, $w_2$, and $l^{-1/2}$ by 
the same factor. 
Consequently, the flow lines, which can be obtained by solving the equation for $dw_2/dw_1$ that results from the flow equations, are 
mapped to one 
another by such a rescaling. For $\rho=w_2/w_1$ small, we find
\be
\frac{dw_2}{dw_1}=\frac{11}{6}\frac{w_2}{w_1}
\ee
to leading order in $\rho$, with solutions $w_2=Bw_1^{11/6}$ for $B$ independent of $w_1$, $w_2$; these exhibit the scaling 
property 
($B$ changes under rescaling).
Note that here the behavior corresponds to the region {\em outside} the separatrix in the flows for $d=6+\eps$, so it is not clear that 
we can even 
get close to the flows that control the tricritical point for $d>6$. 

The crossover to the BR theory occurs when $|r(l_0)|=1$. The flows for $w(l)$ in the unreduced theory for $d=6$ are logarithmic, $w(l)
\propto l^{-1/2}$ 
at weak coupling. For $\rho$, again $\rho(l_0)\sim w(l_0)^2$, so here $\rho(l_0)\sim l_0^{-1}$. To determine the value of $B$ for the 
relevant flow, 
we use $\rho=Bw_1^{5/6}$ at small $\rho$, giving $B\sim l_0^{-7/12}$. This is small at weak field, however for flow lines that start 
near the origin in 
the $w_1$--$w_2$ plane, the smaller the value of  $B$, the larger the coupling $w_1$ at which the flow reaches $\rho=1$. We conclude 
that for weak field 
(large $l_0$), the flows pass out of the domain of validity of perturbation theory before reaching $\rho=1$. Hence our approach breaks 
down,
and we cannot evaluate (at least, not with the present method)  the dependence of, for example, the size of the jump in $q_1$ in terms 
of $h$, 
even though we expect it to be small in the limit $h\to0$. For $d<6$, this effect only becomes worse. It is possible that general methods 
for 
fluctuation-driven first-order transitions \cite{cw,rud,ja} could be useful here.

%%%%%%%%%%%%%%%%%%%%%%%%%%%%%%%%%%%%%
\subsection{Full RSB at lower $T$}
\label{frsb}

For $d$ of order $6+\eps$ or below, we find a transition from $1$-RSB to full RSB as temperature is lowered further below the line 
$T_c(h)$, which occurs
because $\widetilde{y}$ changes sign during the RG flow.  If we work near $T_c$, where Landau-Ginzburg theory is valid, then at weak 
field $\widetilde{y}$ 
(or rather its analog in the unreduced theory) is positive. But we have seen that when the crossover to the BR theory occurs, initially $
\rho<1$, and when 
$\rho(l)$ is large enough, the RG flows drive $\widetilde{y}$ to negative values; this occurs rapidly (over a range of $l$ of order $1$). 
For $d=6+\eps$, this occurs
only at sufficiently large $h$, and the resulting transition line intersects the AT line in the Lifshitz-type point $\cal L$, the location of 
which was already discussed. 
For $d\leq 6$, it occurs right after the crossover. This will occur on a line essentially given by the crossover at which $|r(l_0)|=1$; this 
line is similar to, 
but below, the AT line or its replacement $T_c(h)$.

In the literature on Potts and $p$-spin mean-field SG models \cite{gks,gardner}, there is a transition from $1$-RSB to full RSB that 
occurs at lower temperature;
it is frequently termed the ``Gardner transition''. 
In these theories, typically the parameters $w_i$, $y_j$ that occur in Landau theory are treated as constants, while $r$, $h$ are varied 
as parameters. 
Hence the transition here seems to occur in a different way than in those theories. 

%%%%%%%%%%%%%%%%%%%%%%%%%%%%%%%%%%%%%%%%%%%%%
\section{Discussion: the frozen phase and the metastate of $1$-RSB}
\label{discuss}

In this Section, we take a broad view and discuss the meaning of both the dynamically frozen phase and the $1$-RSB phase, including 
the metastate.
Some issues are uncovered that are common to these two phases and suggest that we lack a full understanding of them in 
short-range systems.

In the mean-field SG models that exhibit a dynamical transition at a temperature $T=T_{\rm d}$ higher than the thermodynamic 
transition temperature $T_c$ \cite{kirk,kw}, what occurs is a breakdown of the ergodicity of the dynamics at and below $T_{\rm d}$ in 
an infinite-size system. 
We note that, at the microscopic level, a theory using Langevin dynamics requires the use of soft spins. In our short-range models, a 
time-dependent 
Landau-Ginzburg theory could be used instead; such a theory would include the static (equilibrium) results of this paper as special cases. 
We expect
that, in a mean-field treatment, a similar dynamical transition would be found at $T_{\rm d}$, but we will not attempt this calculation in 
the present paper. 
Instead, we turn to
the implications for an equilibrium description. A loss of ergodicity in dynamics implies that the configuration space can be broken into a 
number of
ergodic components; by definition, if the (infinite) system were started in a configuration in one of these components, it would never find 
its way into a 
distinct component after any finite time. In a short-range system, it is natural to associate each of these ergodic components 
with a pure state, that is, an extremal Gibbs state of the infinite system (see e.g.\ Refs.\ \cite{ns96b,ns_rev,read14} and references 
therein). 
In the case of the dynamical transition, it appears \cite{kirk,kw} that the entropy of these ergodic components or pure states (calculated 
from 
the measure giving the decomposition of the Gibbs state into pure states), is extensive, that is, the entropy (when defined somehow for 
a finite ``window'' or subregion of the infinite system; the details of this need not concern us now) per unit volume is positive. 

Consequently, if one takes the frozen phase of Landau theory seriously, the theory presented here for low-dimensional SGs appears to 
predict that in the 
regime $T_c<T<T_{\rm d}$, the Gibbs state of the infinite system decomposes into pure states with the associated ``configuration'' 
entropy 
being extensive. For $T<T_c$  in the $1$-RSB state (thus, for $T$ not too low), it is believed that the pure states are characterized by 
random free 
energies that are independent and exponentially distributed, such that the system is condensed into a countable number of pure states 
\cite{mpv_book}; 
hence the entropy of the distribution given by the set of corresponding normalized Boltzmann weights is of order one only. Such 
behavior for $T$ above 
and below $T_c$ (but below $T_{\rm d}$ and above some lower transition temperature) closely resembles the REM \cite{derr,gm}, 
except that energies of the latter are replaced here by free energies, and each individual spin configuration is replaced by a pure state.

There may be legitimate concerns about whether such a picture is truly possible in a short-range, finite-dimensional model. We want to 
argue that this is no more
of an issue for the dynamically-frozen phase than for the $1$-RSB phase, and (with one reservation) no more than for RSB in general. 
(We will leave aside dynamics;
there the issue is that, in a short-range system, a diverging relaxation time on approaching some $T_{\rm d}$ should be accompanied by 
a diverging length scale 
 \cite{ms}, 
but this does not seem to occur at the dynamical transition of Ref.\ \cite{kirk}, or here, on approaching $T_{\rm d}$ from above.) The 
way that a 
many(-pure)-state picture could break down in a short-range model, when it exists in an infinite range model (or mean field description), 
is that it may be possible
to make a droplet, whose interior is essentially in one pure state, say $a$,  as an excitation of another pure state, say $b$. If the 
increase in free energy
on exciting the droplet does not diverge when the size of the droplet goes to infinity, then such droplets will appear thermally on all 
scales, and the distinction 
between the pure states $a$ and $b$ will be lost. If instead the droplet free energy does diverge (at least, if it diverges fast enough), 
then there will only 
be some density of finite size droplets, and this does not destroy pure state $b$. In a dynamical picture, to get from pure state or 
component $a$ 
to $b$ requires thermal activation of a similar droplet; if the droplet free energy diverges with its size then the probability of going from 
$a$ to $b$
in a finite time will be zero, and such a divergence is a necessary condition for the existence of many ergodic components or pure states. 
This is the same issue 
in SG theory that has remained an unresolved controversy for many years, with proponents of RSB and of the scaling-droplet theories 
on the two sides (the 
answer to the question may depend on the dimension of space, but we leave this implicit). 

The reservation in the present case concerns the extensive entropy of the pure states in the dynamically-frozen phase. It can be shown 
rigorously
that, for a given short-range Hamiltonian and at a given temperature, any Gibbs state, and in particular all the pure states, must have 
the same 
free energy density \cite{vH}; by extending the argument, they have the same density of total entropy also. (The former can also be 
seen heuristically. 
A difference in free energy density between two Gibbs states implies that a droplet of the 
phase with lower free-energy density can be created in the higher one; the probability of such a droplet goes to one as it is made 
arbitrarily large.) 
If we define the entropy of the pure state decomposition as seen in a region of volume $W^d$ to be that of the Gibbs state minus the 
average of those 
of the pure states, then it follows that the entropy of the pure state decomposition cannot be extensive, that is $\sim W^d$ \cite{vH}. 
Indeed, we show in 
Appendix \ref{app} that this difference is the mutual information between the spins in the region and the pure states, and that it is 
bounded above
by the surface area $\sim W^{d-1}$ of the finite region. It should be noted that this does not imply that the decomposition into pure 
states is trivial; it appears that  
mutual information of order $W^{d-\zeta'}$ with $\zeta'\geq 1$ cannot be ruled out {\it a priori}. Alternatively, it may be that the frozen 
region is actually ergodic in the short-range cases, but with timescales that diverge as $T\to T_c^+$ (not at $T_{\rm d}$); this view is 
widely held in 
the RFOT community, beginning from Ref.\ \cite{ktw}. We also note that the scaling-droplet theory has been argued \cite{fh_dyn} 
to produce large relaxation times at $h>0$ and $T<T_c(0)$; the difference is the absence of a thermodynamic transition in the latter 
theory at $h>0$ \cite{bm1,fh}. 

Next we turn to the metastate of the SG. The metastate, introduced in SG theory by Newman and Stein \cite{ns96b} (NS), can 
characterize the dependence 
of the Gibbs state
in a finite region (or ``window'') on the finite size of the system, or alternatively on the disorder far away [the latter giving the 
Aizenman-Wehr \cite{aw} (AW) 
 metastate;
the two constructions are known to be equivalent at least in some cases]. Formally, a metastate is a probability distribution on infinite-
size Gibbs states,
for given disorder. An earlier paper \cite{read14} introduced a method for studying the AW metastate using RSB. Here we will apply this 
to the case of $1$-RSB, 
and allow the possibility of a magnetic field.  

Using the AW metastate, we can define the average of the Gibbs state with respect to the metastate for given disorder.
For correlation functions, this corresponds to averaging the correlation function over the metastate, for given disorder; we say that this 
gives a correlator
in the ``metastate-averaged state'' (MAS). To obtain this from finite size, we use several copies of a finite-dimensional system of size 
$L$ in a box with free 
boundaries; in an outer region at distance $>R$ from the origin ($R<L$) these have independent samples of the disorder (the bonds and 
the magnetic fields, 
if the latter are random), while in the inner region at distance $<R$ they are identical. For the case of a square of a correlator in the 
MAS, we can use 
this construction; taking the average over the disorder in the outer region corresponds to using the MAS. Then we square and average 
over the disorder in the 
inner region, and take $L$, $R\to\infty$ \cite{read14}. 

In the present case, with a magnetic field, the MAS correlation function of interest can be represented by first considering (in finite size) 
the average
\be
C_{ij}=\left[\left(\left[\langle s_i s_j\rangle\right]_>-\left[\langle s_i \rangle\right]_>\left[\langle s_j\rangle\right]_>\right)^2\right]_<.
\ee
Here $[\cdots]$ stands for a disorder average, while $\langle\cdots\rangle$ is a thermal average, initially in finite size. The square 
brackets with subscripts $>$ and $<$ 
denote averages over only the disorder in the outer and inner regions, respectively. On taking the limits $L\to\infty$, then $R\to\infty$, 
this gives the correlation 
function of interest. 

This correlator resembles that in the definition of the SG susceptibility, namely
\be
\chi_{ij}=\left[\left(\langle s_i s_j\rangle-\langle s_i \rangle\langle s_j\rangle\right)^2\right],
\ee
which however differs in that all disorder is averaged over at the end. The SG correlator $\chi_{ij}$ tends to zero at large distance 
$|\bx_i-\bx_j|$ (where for any $i$,
site $i$ is located at $\bx_i$ on the lattice) whenever the thermal average 
$\langle\cdots\rangle$ is taken in a pure state, such as at high temperature. When the Gibbs state is not pure, $\chi_{ij}$ usually tends 
to a constant, which using RSB
can be expressed in terms of integrals of $q(x)$; hence the SG susceptibility, which is the sum of $\chi_{ij}$ with respect to $j$ (say) 
running over all space, diverges 
like the volume in any SG ordered phase, for example below the AT line according to the standard full RSB picture. In the present case, 
in the frozen region 
$T_c<T<T_{\rm d}$, $\chi_{ij}$ tends to zero at large distance, because in integrals the resulting $q(x)$ is indistinguishable from the 
constant ($=Q$) that occurs 
in the high-temperature region, and which cancels as for a pure state. 

The idea of the correlator $C_{ij}$ is that, if there is no dependence of the two-point spin correlation on the distant disorder, then it 
becomes the same as the
SG correlator $\chi_{ij}$, and so goes to a constant in the SG phase. (This will occur if the metastate is trivial, that is if it is supported on 
a single Gibbs state.)
But if there is such dependence that remains in the limit $R\to\infty$ then the metastate is non-trivial. The MAS is itself a Gibbs state, 
and so can be decomposed into 
pure states, and for a non-trivial metastate it is expected that there are many pure states in that decomposition, and consequently 
$C_{ij}$ will decay as 
$|\bx_i-\bx_j|\to\infty$; RSB theory predicts \cite{read14} that in fact it will tend to zero in this limit. 

In Ref.\ \cite{read14}, it was argued that the use of distinct disorder in two samples in the outer region breaks the symmetry among the 
replicas; the 
replicas are divided into
groups that in effect arise from different copies in the outer region. Then a correlation function of spins well in the interior of the inner 
region as $R\to\infty$ can
 be calculated by taking spins with replica indices in different groups. 
In the replica formalism, using $Q_{\alpha\beta}(\bx_i)\sim s_i^\alpha s_i^\beta$, $C_{ij}$ becomes the correlation function
\bea
C_{ij}&=&\langle\!\langle \widetilde Q_{\alpha\beta}(\bx_i)\widetilde{Q}_{\alpha\beta}(\bx_j)\rangle\!\rangle
           -2\langle\!\langle \widetilde Q_{\alpha\beta}(\bx_i)\widetilde{Q}_{\alpha\gamma}(\bx_j)\rangle\!\rangle\non\\
          &&{} +\langle\!\langle \widetilde Q_{\alpha\beta}(\bx_i)\widetilde{Q}_{\gamma\delta}(\bx_j)\rangle\!\rangle,
\eea
where $\langle\!\langle\cdots\rangle\!\rangle$ stands for an average in the Landau-Ginzburg theory, and each of $\alpha$, \ldots, $
\delta$ belongs to a distinct group. 
Here we subtracted off the replica symmetric part $Q$ (which in the BR theory is non-fluctuating) to arrive at the fields 
$\widetilde{Q}_{\alpha\beta}$; the $Q$ terms cancel. It was argued in Ref.\ \cite{read14} that the components in distinct groups 
correspond to
the outermost blocks in the RSB form (before $n\to0$). Thus the expectation $\langle\!\langle\widetilde{Q}_{\alpha\beta}\rangle\!
\rangle$ is equal to $q_0$, 
which is zero above $T_c$, and negative below. These expectations again cancel, so we can replace $\widetilde{Q}_{\alpha\beta}$ by 
$\delta\widetilde{Q}_{\alpha\beta}=\widetilde{Q}_{\alpha\beta}- \langle\!\langle\widetilde{Q}_{\alpha\beta}\rangle\!\rangle$, which 
are the fluctuations 
around the ordered phase. 

In the cases studied here, which involve at most $1$-RSB, we know that at the Gaussian level, all modes have positive mass-squared, 
except at some of the 
transitions. Here we discuss only the non-critical properties.  It now follows, without detailed calculation, that the correlator $C_{ij}$ 
decays exponentially to zero as 
$|\bx_i-\bx_j|\to\infty$ in all such cases. Note that in the SG phase, this is distinct from the SG correlator $\chi_{ij}$, which goes to a 
constant. It means that there 
are many  pure states in the MAS; the metastate is highly non-trivial. 

In Ref.\ \cite{read14}, it was argued that in a SG phase, we would have
\be
C_{ij}\sim \frac{1}{|\bx_i-\bx_j|^{d-\zeta}},
\ee
where $\zeta$ is universal and gives information about the metastate; $\zeta\leq d$. For a trivial metastate, $\zeta=d$. (See also 
Refs.\ \cite{wf,wy,billoire}.) 
 Exponential decay should be considered equivalent to $\zeta$ being at its other bound $\zeta=0$, as in both cases the integral of 
$C_{ij}$ over $\bx_i-\bx_j$ 
just converges (up to logarithms). It was further 
argued that, if we examine the MAS only in a window of size $W$, the logarithm of the number of pure states in the MAS that can be 
distinguished within 
the window (see Appendix \ref{app} for this notion and an improved definition as mutual information) scales as $W^{d-\zeta'}$ (when 
nonzero), and that 
$\zeta'=\zeta$ (we hope 
that no confusion will arise from using the same symbol $\zeta'$ as for the similar exponent in the frozen phase; we do not mean that the 
two values must necessarily 
be equal). $\zeta'=0$ would mean an extensive entropy of pure states.  

In Ref.\ \cite{read14}, it was also argued that in short-range models there is a lower bound $\zeta\geq1$. This bound, which 
strengthens the result 
that the entropy of pure states must be sub-extensive, was obtained from the obvious fact that at zero temperature, there cannot be 
more than 
$2^{{\cal O}(W^{d-1})}$ ground states in a window of size $W$, so $\zeta'\geq 1$, together with the belief that the full RSB phase can 
be continued 
to zero temperature, with the exponent $\zeta'$ unchanged, and then $\zeta'=\zeta$ implies that the same bound arises for full RSB. As 
mentioned 
already, in Appendix \ref{app} we have proved such a bound on the mutual information directly at any temperature; hence, when $
\zeta'$ is defined, 
$\zeta'\geq 1$. The present 
result that $\zeta=0$ does not satisfy the bound if $\zeta=\zeta'$, and hence there is an inconsistency somewhere in the arguments. 
Perhaps 
either $\zeta'\neq \zeta$, or an improved calculation of $C_{ij}$ would produce a larger $\zeta$. We note that if the MAS has trivial 
decomposition 
into pure states, then exponential decay to zero of $C_{ij}$ is possible (e.g.\ in the high temperature phase); then $\zeta=0$ but $
\zeta'$ is not defined 
when the mutual information is zero. 

In the frozen phase $T_c<T<T_{\rm d}$, $C_{ij}$ decays exponentially, and so does $\chi_{ij}$. This strongly suggests that the 
metastate is 
trivial in this region, even though the Gibbs state may contain a large number of pure states. (This form has been discussed as 
a possible scenario \cite{ns96b,read14}.) Indeed, we can show \cite{hr} in leading order within Landau-Ginzburg theory that in this case 
$C_{ij}=\chi_{ij}$, as for a trivial metastate, and that both are equal to $\chi_{ij}$ of the paramagnetic (replica symmetric) phase, 
given in Fourier space by appropriate components of expression (\ref{rs_prop}). (If the metastate is trivial then the MAS is the same as 
the Gibbs state,
and the use of the same notation $\zeta'$ is justified.) It is tempting to think that 
on passing through $T_c$, the latter pure states become those involved in the non-trivial metastate 
of the $1$-RSB phase, just as in the REM picture of the transition. (We should caution that the idea that the pure states are the 
``same'' should 
not be taken too literally, due to the expected ``temperature chaos'' \cite{bm87,fh}, the sensitivity at large length scales of the Gibbs 
(and pure) states to small changes in temperature.) In fact, the MAS in the $1$-RSB region appears to have a very similar structure to 
the 
Gibbs state in the frozen phase, and if the metastate is trivial in the latter, then the MAS 
could be virtually unchanged on passing through the transition to $1$-RSB, with the same values of both $\zeta$ and $\zeta'$. 

We also mention here a simpler criterion for non-triviality of the metastate, given in Ref.\ \cite{read14}. If
\be
\left[\langle s_i\rangle^2\right]-\left[\left[\langle s_i\rangle\right]_>^2\right]_< >0
\label{MASs}
\ee
(in the limit), then the metastate is non-trivial (the converse may also hold, but that is not completely clear). This means that the 
conditional 
variance of $\langle s_i\rangle$ due to the distant disorder (for given disorder in the inner region), averaged over disorder in the inner 
region, is 
nonzero (in the limit). In terms of RSB, it reduces to 
\be
\int_0^1\!q(x)\,dx-q(0) >0,
\label{qcrit}
\ee
and in terms of $\qt(x)$ becomes $\qt(0)<0$; thus for the one-step cases in this paper, this is $q_0 < 0$. This immediately 
gives all the non-triviality results above, however it does not yield the additional information provided by $\zeta$.
It is interesting to note that the left-hand side is of order $\tau-\tau_c$ (to leading order) in all cases
considered in Landau theory here; thus this order parameter for a non-trivial metastate has exponent $\beta=1$. In the case of the 
discontinuous transition,
there are no critical fluctuations, so the same should hold even in dimensions $d<6$ at that transition. 

To sum up the arguments presented in this section, we find that once one considers the metastate or MAS, there are 
consistency issues not only for the frozen phase, but also similar ones for the $1$-RSB phase, in relation to Ref.\ \cite{read14}. These 
issues with these phases 
arise not only for the AT line in the Ising 2-spin interaction model considered in this paper, but also for other models in short-range 
finite-dimensional cases (e.g.\ Potts, $p$-spin models, and so on) that may possess similar phases. We are not able to 
resolve these issues at present.

%%%%%%%%%%%%%%%%%%%%%%%%%%%%%%%%%%%%%%%%
\section{Conclusion}
\label{concl}

In this work, we considered the BR reduced Landau-Ginzburg theory for a SG, extended by the addition of quartic terms. This theory
is the most general one for a RSB transition in the absence of any symmetry. Here it arose because we considered the AT line for 
an Ising SG in a magnetic field, or for other SGs in random isotropic magnetic fields. This theory is ``reduced'' in that it contains only
the so-called replicon modes, which are the ones involved in RSB \cite{par79}. Other fluctuating modes do not generically become 
massless at the same
point as the replicons, in the absence of symmetry; hence they can be neglected or integrated out to leave the reduced theory. 
Consequently we believe that this theory is of widespread applicability. The case of a SG with no symmetry is a most basic case; SGs with 
additional
symmetry, which usually (except for the Ising case) have additional indices (other than replica indices $\alpha$, $\beta$) on the 
$Q_{\alpha\beta}$ 
field on which the symmetry operations act, are more specialized. 

Based on a renormalization-group (RG) analysis of this theory, we showed that for the AT line at $d$ close to $6$, fluctuations cause the 
effective 
value of the combination $\widetilde{y}$ of the quartic couplings to become negative, signaling a continuous transition to a one-step RSB 
($1$-RSB) 
phase at weak (but not extremely weak) magnetic field. At slightly higher magnetic field, there is a tricritical point $\cT$, and the 
transition becomes 
a quasi-first-order transition to $1$-RSB beyond that point. We suggest that this latter transition, in particular its quasi-first-order 
character, could 
persist for any non-zero field when $d\leq6$, giving a possible resolution of the long-standing problem first posed by BR. We emphasize 
that our 
results for $d>6$ and magnetic fields that are not too large are well controlled within the perturbative RG treatment for the Landau-
Ginzburg theoy, 
similar to an epsilon expansion; of course, we cannot rule out the existence of non-perturbative effects that invalidate perturbation 
theory itself, but such 
effects have not been discovered for the present theory to our knowledge. We also mention that results of the same form apply in other 
models, 
including the power-law one-dimensional model \cite{kas} in a magnetic field, in the region $\sigma=2/3-\eps$ ($\eps>0$) that 
corresponds to $d=6+\eps$. 

We want to comment here on the possible implications for other techniques for studying the AT line, such as Monte Carlo simulation. The 
most common 
method of searching for a transition
in a magnetic field has been to look for a divergence of the SG susceptibility, or of the related correlation length. Unfortunately, these 
methods can only detect
a second-order transition, and negative results at $d<6$ (and in the corresponding regime in power-law models) have been interpreted 
as meaning that 
there may be no transition. The problem should be studied with methods that can detect a (quasi-)first-order transition. To do so, we 
suggest use of some 
of the diagnostics mentioned near the end of Sec.\ \ref{discuss}. These were (i) the divergence (as the volume) of the SG susceptibility 
in both the $1$-RSB 
and full RSB phases; (ii) the MAS correlation function $C_{ij}$ and its exponent $\zeta$ ($\zeta<d$ means a non-trivial metastate, which 
does not occur 
in the high-temperature phase), or (iii) perhaps most simply, the single-spin average as in ineq.\ (\ref{MASs}), which again would signal 
that the metastate 
is non-trivial. 

We also comment that Ref.\ \cite{janus1} found some evidence of a dynamical transition in three dimensions and near where the AT line 
would be.
They also found evidence of an ordered phase using the criterion of the form of ineq.\ (\ref{qcrit}), obtained from Monte Carlo 
evolution. These results 
seem consistent with aspects of our findings (see Ref.\ \cite{kirk} and Sec.\ \ref{discuss} above). Other results for a range of 
dimensions were obtained 
from high-temperature series \cite{singh} that studied the SG susceptibility; evidence of an AT line was found for $d\geq 6$, but for 
$d=5$ most Pade 
approximants showed no divergence of $\sum_j\chi_{ij}$. Clearly these results agree with ours to some degree. The fact that there 
appeared to be a 
second-order transition for $d=6$ could be a consequence of it being the borderline case; the quasi-first-order behavior might be very 
weak there, 
and the SG susceptibility, though finite, could be large (at the fields studied). 

In this work we used the RG approach in a simple way. To consider the possible quasi-first-order transition for $d\leq6$, more powerful 
RG methods are required. 
We hope to return to this elsewhere.

\acknowledgments

NR is grateful to B. Chakraborty, P. Charbonneau, P. Goldbart, C. Newman, D. Stein, M. Moore, D. Fisher, G. Biroli, and L.-P. Arguin for 
emails
and discussions, and M. Moore for a previous collaboration. 
He is also grateful to C. O'Hern for organizing the 4th International Conference on Packing Problems at Yale University in June 2019,
which helped stimulate the ideas contained herein. Both authors acknowledge the support of NSF grant no.\ DMR-1724923.

%%%%%%%%%%%%%%%%%%%%%%%%%%%%%%%%%%%
\begin{appendix}
\section{Entropy of pure-state decomposition as mutual information, and a bound}
\label{app}

In this Appendix, we introduce a definition for the entropy (associated with a finite window) of the decomposition of an (infinite-size) 
Gibbs state into pure states, valid in short-range systems of spins, each of which takes a finite number of states (for example, Ising 
spins). 
We interpret it as the mutual information between the spins in the window and the pure states, derive a bound on it, and consider some 
examples. 
We also critique the idea of counting the pure states that are ``distinguishable within the window'' 
(that is, those that differ by more than some amount). We have in mind a SG, though we will not need to average over disorder, and the 
results are very general. We will assume the system is a hypercubic lattice in $d$ dimensions with nearest-neighbor pair-wise 
interactions, though 
the arguments can easily be generalized considerably. The main result is easy to prove, but we are not aware of a discussion of it within 
the 
statistical mechanics literature. These are foundational results about Gibbs states.

We define a window $\Lambda$ that is a hypercube of side $W$, with edges parallel to those of the 
hypercubic lattice; we also view $\Lambda$ as a set of sites 
$\Lambda=\{i\in\Lambda\}$ in the lattice. The set of lattice sites on the surface of $\Lambda$, the spins on which have an interaction 
with at least one spin 
outside $\Lambda$, is denoted $\partial\Lambda$. Spins at sites in the interior $\Lambda^o=\Lambda-\partial\Lambda$ of $\Lambda$ 
interact 
only with others in $\Lambda$. $\Lambda^c$ is the complement of $\Lambda$, $\Lambda^c={\bf Z}^d-\Lambda$. As before, we 
denote the 
spins by $s_i$, and write $s_\Lambda$ for $(s_i)_{i\in \Lambda}$, and similarly for
$s_{\partial \Lambda}$ and $s_{\Lambda^o}$. 

We assume that the system possesses Gibbs states. 
A Gibbs state $\Gamma$ with temperature $T\geq 0$ is a probability measure on the spins that satisfies the condition that, for any 
window $\Lambda$, 
the conditional probability distribution $\Gamma(s_\Lambda|s_{\Lambda^c})$ (i.e.\ conditioned on the spins outside $\Lambda$) is the 
usual 
Gibbs distribution $\propto e^{-H(s_\Lambda,s_{\Lambda^c})/T}$, where the fixed spins $s_{\Lambda^c}$ are treated as a boundary 
condition. 
A Gibbs state has a unique decomposition into pure Gibbs states, $\Gamma=\sum_\alpha w_\alpha\Gamma_\alpha$, where $w_\alpha$ 
are nonnegative
and $\sum_\alpha w_\alpha=1$ (see e.g.\ Refs.\ \cite{ns96b,ns_rev,read14} and references therein; the assumption that the $
\Gamma_\alpha$ are pure 
will not be used). (We write a sum for ease of writing, but it may be that it should be an integral employing some measure on an 
uncountable set of $\alpha$ instead, or a sum plus an integral. Such a change can be made throughout, and causes few difficulties, 
except 
in an alternative approach that we discuss at the end.) 

It will be useful to view $w_\alpha\Gamma_\alpha$ as a joint probability measure for both spin configurations and pure states. We will be 
most interested in its marginal distribution $w_\alpha\Gamma_\alpha(s_\Lambda)$ (i.e.\ ignoring spins outside $\Lambda$) which gives 
the 
joint probability distribution for the pair of random variables ${\cal S}_{\Lambda}$, ${\cal A}$. Its marginal distribution on spins alone is 
$\Gamma(s_\Lambda)$, and on pure states alone is $w_\alpha$; the conditional probability that ${\cal S}_\Lambda=s_\Lambda$ given 
that ${\cal A}=\alpha$ is $\Gamma_\alpha(s_\Lambda)$.   

We can now give our proposed definition for the entropy for a finite window of the decomposition into pure states. As 
${\cal S}_\Lambda$ 
takes a finite set of values, it has well-defined entropy using the marginal $\Gamma$:
\be
S({\cal S}_\Lambda)=-\sum_{s_\Lambda}\Gamma(s_\Lambda)\ln \Gamma(s_\Lambda).
\ee
We use a similar definition for $S_\alpha({\cal S}_\Lambda)$, with $\Gamma_\alpha$ in place of $\Gamma$. We can then subtract the 
average of $S_\alpha$ (with respect to the probabilities $w_\alpha$) from $S$. This should isolate the entropy ``due to'' forming the 
mixture using the weights $w_\alpha$, and so correspond to the entropy of the decomposition, relativized to the finite window 
$\Lambda$. 
(Similar definitions, though not relativized to a finite window, were given in Refs.\ \cite{vH,palmer}.) Thus our proposal is to use
\be
S({\cal S}_\Lambda)-\sum_\alpha w_\alpha S_\alpha({\cal S}_\Lambda)=I({\cal S}_\Lambda;{\cal A}).
\ee
Here we have identified the difference as the 
{\em mutual information} $I({\cal S}_\Lambda;{\cal A})$ of the spin configuration ${\cal S}_\Lambda$ and the pure state ${\cal A}$. 
The
definition is based on the joint probability measure $w_\alpha\Gamma_\alpha(s_\Lambda)$. The standard way to write the definition of 
the mutual information of two random variables $A$, $B$ with joint distribution $p(A=a,B=b)$ is  \cite{ct_book}
\bea
I(A;B)&=&S(A)+S(B)-S(A,B)\\
&=&S(A)-S(A|B);
\eea
here we used physics notation $S$ for the Shannon entropy, in place of the more standard notation $H$ used in information theory. 
Thus we 
have $A={\cal S}_\Lambda$, $B={\cal A}$, and we can identify $\sum_\alpha w_\alpha S_\alpha({\cal S}_\Lambda)$ as the conditional 
entropy 
$S({\cal S}_\Lambda|{\cal A})$. (If $B$ is continuous, but not $A$, then some of the terms containing $B$ in 
these expressions suffer the usual ambiguity resulting from the need to define a reference measure for the integral over the variable in 
terms of 
a density, the logarithm of which is taken to define the entropy. This ambiguity cancels in the conditional entropy $S(A|B)$ and in the 
mutual information 
$I(A;B)$; in particular, it does so in our case.) The mutual information $I({\cal S}_\Lambda;{\cal A})$ can be viewed as a measure of 
how much information 
we obtain about which pure state we are in from the spin configuration in the window. A standard application of Jensen's inequality 
implies that 
$I(S_\Lambda;{\cal A})\geq 0$.

We now observe that we can break $S$ into two terms, by conditioning on the spins in $\partial \Lambda$:
\be
S({\cal S}_\Lambda)=-\sum_{s_\Lambda}\Gamma(s_\Lambda)\ln \Gamma(s_{\partial \Lambda})-\sum_{s_\Lambda}\Gamma(s_\Lambda)
\ln \Gamma(s_{\Lambda^o}|s_{\partial\Lambda}).
\ee
The first term is the entropy $S({\cal S}_{\partial\Lambda})$ of ${\cal S}_{\partial \Lambda}$, the spin configuration on the boundary 
only, while 
the second is the conditional entropy 
$S({\cal S}_{\Lambda^o}|{\cal S}_{\partial\Lambda})$ of the 
interior given the boundary. But because the interactions are between nearest neighbors only, the conditional probabilities 
$\Gamma(s_{\Lambda^o}|s_{\partial\Lambda})$ are determined
by the definition of a Gibbs state (here for the window $\Lambda^o$ rather than $\Lambda$), and consequently are independent of the 
Gibbs state 
$\Gamma$ or $\Gamma_\alpha$ with which we started. We can do the same with $S_\alpha$, and as 
\bea
\Gamma(s_\Lambda)=\Gamma(s_{\Lambda^o},s_{\partial\Lambda})&=&\Gamma(s_{\Lambda^o}|s_{\partial\Lambda})
\Gamma(s_{\partial\Lambda})\\
&=&\sum_\alpha w_\alpha\Gamma_\alpha(s_{\partial\Gamma})\Gamma(s_{\Lambda^o}|s_{\partial\Lambda}), 
\eea
we see that, in the difference, {\em the interior (conditional entropy) terms cancel}. We are left with
\be
I({\cal S}_\Lambda;{\cal A})=S({\cal S}_{\partial\Lambda})-\sum_\alpha w_\alpha S_\alpha({\cal S}_{\partial\Lambda})=
I({\cal S}_{\partial\Lambda};{\cal A}).
\ee
(The general case of this result is contained in the proof of Thm.\ 2.8.1 in Ref.\ \cite{ct_book}.)
Each term in the middle expression is an entropy on the boundary, and is non-negative, 
so we immediately obtain (for the case of Ising spins)
\bea
I({\cal S}_\Lambda;{\cal A}) &\leq& S({\cal S}_{\partial\Lambda}) \label{mi_ineq1}\\
&\leq&  |\partial \Lambda|\ln 2.
\label{mi_ineq2}
\eea
Thus at any $T\geq 0$ the mutual information is at most of order the surface area $\propto W^{d-1}$ of $\Lambda$. This is the main 
result. 

We now make a few remarks on why we believe this is a reasonable way in which to define the finite-volume entropy
of the decomposition into pure states, and make some additional points. First, suppose that we have probabilities (not necessarily Gibbs 
states) $\Gamma(s_\Lambda)
=\sum_\alpha w_\alpha \Gamma_\alpha(s_\Lambda)$, where the probabilities $\Gamma_\alpha$ for distinct $\alpha$ are nonzero on 
disjoint sets of configurations, and these subsets are indexed by $\alpha$. Then the mutual information is 
\be
I({\cal S}_\Lambda;{\cal A}) = - \sum_\alpha w_\alpha \ln w_\alpha,
\ee
which indeed is the entropy (or ``complexity'' \cite{palmer}) $S({\cal A})$ of $\cal A$. This occurs because the conditional entropy 
$S({\cal A}|{\cal S}_\Lambda)=0$ in this case.
 
Of course, such a partition of the configurations does not usually occur in Gibbs states in practice. For Gibbs states in general, we obtain 
instead
\be
I({\cal S}_\Lambda;{\cal A}) =  \sum_{s_{\partial\Lambda},\alpha} w_\alpha\Gamma_\alpha(s_{\partial\Lambda})
\ln\frac{\Gamma_\alpha(s_{\partial \Lambda})}
{\sum_{\alpha'} w_{\alpha'} \Gamma_{\alpha'}(s_{\partial \Lambda})}.
\ee
We may make some observations about this expression. One is that distinct pure states may become identical when restricted to their 
marginals 
$\Gamma_\alpha(s_{\partial\Lambda})$. Then we may as well define the random variable ${\cal A}_\Lambda$, with values $
\alpha_\Lambda$, such that
the (marginal) distribution for ${\cal A}_\Lambda$ is
\be
W_\Lambda(\alpha_\Lambda=\alpha)=\sum_{\alpha': \forall s_\Lambda, \Gamma_{\alpha'}(s_\Lambda)
=\Gamma_{\alpha}(s_\Lambda)} w_{\alpha'},
\ee
that is, the weights of pure states that are identical in $\Lambda$ have been summed.  [As any $\Gamma_\alpha(s_{\partial\Lambda})$ 
is equivalent to a collection of $2^{|\partial\Lambda|}$ probabilities that sum 
to $1$, they form a space (a simplex) of dimension $2^{|\partial\Lambda|}-1$, and a probability measure on that space is sufficient to 
characterize 
$W_\Lambda(\alpha_\Lambda)$.] For discrete $\alpha_\Lambda$, another bound on the mutual information becomes tighter after this is 
done, namely 
\be
I({\cal S}_\Lambda;{\cal A})=I({\cal S}_\Lambda;{\cal A}_\Lambda) \leq S({\cal A}_\Lambda),
\label{mi_ineq3}
\ee
which follows from non-negativity of $S({\cal A}_\Lambda|{\cal S}_\Lambda)$ for discrete $\alpha_\Lambda$. The right-hand side 
is the definition we might have anticipated for the entropy of the pure-state decomposition (similar to Refs.\ \cite{vH,palmer}), at least 
when the $\alpha_\Lambda$
are discrete. In the case that discreteness of $\alpha_\Lambda$ holds for $T>0$, such a definition is reasonable; but we note again that 
even then 
it is not clear in general why, in addition, the conditional entropy $S({\cal A}_\Lambda|{\cal S}_\Lambda)$ should be zero, and hence the 
definitions
would not be equivalent.
 
We can also see that use of the mutual information is consistent with the entropy of a distribution on ground states, for which the upper 
bound by 
the surface area is obvious. At zero temperature, in a SG with a continuous distribution of bonds, a pure state is a ground state, which is 
a configuration 
the energy of which does not decrease on changing the values of 
any finite set of spins. Then the values $\alpha_\Lambda$ correspond one-to-one 
with spin configurations $s_{\partial\Lambda}$ (these determine the spins in the interior, because the configuration is a ground state). 
[Thus we now have the case mentioned earlier, of disjoint support of each $\Gamma_\alpha(s_{\partial\Lambda})$.] For the mutual 
information
of perfectly-correlated variables such as these, the upper bounds are both saturated, and one has $I({\cal S}_{\partial\Lambda};{\cal 
A}_\Lambda)=
S({\cal S}_{\partial\Lambda})=S({\cal A}_\Lambda)= - \sum_{\alpha_\Lambda} W_\Lambda(\alpha_\Lambda) \ln 
W_\Lambda(\alpha_\Lambda)$, 
as for the case of disjoint supports. It is clear that this is the correct result. 

For any $T\geq 0$, if the $\alpha$s form a countable discrete set, then so do the $\alpha_\Lambda$. If we consider the dependence on 
the size $W$ of $\Lambda$, 
then the first inequalities (\ref{mi_ineq1},\ref{mi_ineq2}) imply that the mutual information goes to zero as $W\to0$. When $\alpha$ are 
discrete, as $W$ increases 
the second upper bound (\ref{mi_ineq3}) becomes $S({\cal A}_\Lambda)\to S({\cal A})$ which, if finite, is eventually tighter than the 
first. For large $W$ we expect
in this case that 
\be
\lim_{W\to\infty}I({\cal S}_\Lambda;{\cal A}) =  S({\cal A});
\ee
this is because in this situation, when $W$ is large, 
knowledge of the spins ${\cal S}_\Lambda$ should determine the pure state ${\cal A}=\alpha$ with high confidence, so $S({\cal A}|{\cal 
S}_\Lambda)\to 0$.
There will be a length scale $W\sim \xi$, say, at which $I({\cal S}_\Lambda;{\cal A})$ crosses over to $S({\cal A})$. In any RSB phase 
with a plateau 
in $q(x)$ [or $\qt(x)$] extending from $x=x_1$ to $1$, the entropy (or complexity) of $\cal A$ within mean-field or Landau theory is 
\cite{derr,gm}
\be
S({\cal A})=\psi(1)-\psi(1-x_1),
\ee
where $\psi$ is the digamma function, and so $S({\cal A})\to\infty$ as $x_1\to 1$. In particular, this occurs in the $1$-RSB phase at the 
discontinuous 
transition as $T\to T_c^-$, and implies that $\xi\to\infty$ at that transition; this diverging length or vanishing mass-square at this 
transition does 
not show up in the conventional analysis of fluctuations about the $1$-RSB solution (i.e.\ in the eigenvalues of the Hessian) to which we 
referred in 
Secs.\ \ref{critical} and \ref{discuss} \cite{ck}. 

In the more general case (for a non-trivial decomposition into pure states and at $T>0$) that the $\alpha_\Lambda$ are not a countable 
discrete set, 
$S({\cal A}_\Lambda)$ suffers from the ambiguity already mentioned, and the right-hand side of the second upper bound is not well-
defined, 
and can even be made negative, so no bound can be obtained. We may try to make $S({\cal A}_\Lambda)$ well defined by 
approximating it by replacing 
$\alpha_\Lambda$ with a discrete set, say $\alpha_r$ for a countable set of $r$, and attaching a weight $w_r$ to each; each $w_r$ 
should 
be the integral of $W(\alpha_\Lambda)$ over a neighborhood of $\alpha_r$, and the neighborhoods should form a partition of the space 
of $\alpha_\Lambda$,
so $\sum_rw_r=1$. Calculating the entropy using the $w_r$s is then a Riemann-sum type of approximation of $S({\cal A}_\Lambda)$. 
This is a refined version 
of the prescription of taking the logarithm of the number of pure states that can be viewed as distinguishable probability distributions $
\Gamma_\alpha(s_\Lambda)$  
(say, differing by at least $\epsilon$ with respect to some metric). We want to point out, first, that the use of the integral over a 
neighborhood eliminates 
the ambiguity resulting from a reference measure in defining the probability density on $\alpha_\Lambda$. However, there are similar 
issues because we had to make a 
choice of how to partition the space of $\alpha_\Lambda$. Ignoring that, we can then ask about the dependence on $\epsilon$. If 
$\Delta$ is the volume of each 
neighborhood in the partition (so, in $\cal N$ dimensions, we might have $\Delta=\epsilon^{\cal N}$), then for the entropy we have
\bea
S&=&{}-\sum_r w_r\ln w_r\\&\simeq&{}-\sum_r W_\Lambda(\alpha_r)  \Delta\ln [ W_\Lambda(\alpha_r) \Delta] \\
&\sim&{} - \ln \Delta -\int d\alpha_\Lambda\, W_\Lambda(\alpha_\Lambda)\ln W_\Lambda(\alpha_\Lambda)
\eea
as $\Delta\to0$, which diverges \cite{ct_book}. Here we treated $W_\Lambda(\alpha_\Lambda)$ as an integrable function, which might 
not be true---it could represent a 
singular measure, but in that case the integral formula for entropy cannot be used anyway. [For a general measure, using the same 
procedure and extracting the 
coefficient of $\ln 1/\epsilon$ (i.e.\ the leading term as $\epsilon\to0$) yields by definition the information (fractal) dimension of the 
measure. The 
information dimension is $\leq {\cal N}$ in general, so $\leq2^{|\partial\Lambda|}-1$ in our case.] In the absence of more detailed 
information about the measure, 
we cannot improve the argument. If the pure states are concentrated in clusters in $\alpha_\Lambda$, then the discrete version of 
$S({\cal A}_\Lambda)$ for a 
well-chosen $\Delta$ or $\epsilon$ might be a good approximation to $I({\cal S}_\Lambda;{\cal A}_\Lambda)$, and this might work at 
either very low temperature 
or very large $W$. In general the attempt to use some $\epsilon$ to distinguish the pure states when calculating $S({\cal A}_\Lambda)$ 
seems to give results 
that will depend on $\epsilon$ in unpleasant ways.  No such issues arose for our definition, which led to $I({\cal S}_\Lambda;{\cal A})$; 
the latter can be
approximated arbitrarily well by discretization \cite{ct_book}.

For spins that take continuous values, for example if each $s_i$ is an $m$-component unit vector ($m>1$), the mutual information 
$I({\cal S}_\Lambda;{\cal A})$ 
is again well defined and non-negative \cite{ct_book}, but the upper bounds we used in ineqs.\ (\ref{mi_ineq1}) and (\ref{mi_ineq2}) 
are no longer valid.

\end{appendix}

%%%%%%%%%%%%%%%%%%%%%%%%%%%%%%%%%%

\end{document}